\documentclass[aps,twocolumn,prd,preprintnumbers,groupedaddress,nofootinbib,amssymb,eqsecnum,epsfig]{revtex4-1}
\usepackage{graphicx}
\usepackage{bm}
\usepackage{amsmath}
\usepackage{color}
\usepackage{amsfonts}
\usepackage[subnum]{cases}

\begin{document}
\preprint{YITP-17-99} 

\newcommand{\newc}{\newcommand}

\newc{\be}{\begin{equation}}
\newc{\ee}{\end{equation}}
\newc{\bea}{\begin{eqnarray*}}
\newc{\eea}{\end{eqnarray*}}
\newc{\D}{\partial}
\newc{\ie}{{\it i.e.} }
\newc{\eg}{{\it e.g.} }
\newc{\etc}{{\it etc.} }
\newc{\etal}{{\it et al.}}
\newcommand{\nn}{\nonumber}
\newc{\ra}{\rightarrow}
\newc{\lra}{\leftrightarrow}
\newc{\lsim}{\buildrel{<}\over{\sim}}
\newc{\gsim}{\buildrel{>}\over{\sim}}
\def\mpl{M_{\rm pl}}
\def\d{\mathrm{d}}
\def\beq{\begin{eqnarray}}
\def\eeq{\end{eqnarray}}
\def\l({\left(}
\def\r){\right)}
\def\v{\upsilon}
\def\calH{{\cal H}}
\newcommand{\blue}{\color{blue}}
\newcommand{\SK}[1]{{\blue #1}}
\newcommand{\remb}[1]{{\bf\blue [#1]}}

\title{Observational signatures of the parametric amplification of gravitational waves \\ 
during reheating after inflation}

\author{
Sachiko Kuroyanagi$^{1}$, 
Chunshan Lin$^{2,3}$,
Misao Sasaki$^{2}$,
Shinji Tsujikawa$^{4}$}

\affiliation{
$^1$Department of Physics, Nagoya University, Chikusa, Nagoya 464-8602, Japan\\
$^2$Yukawa Institute for Theoretical Physics, Kyoto University, 606-8502, Kyoto, Japan\\
$^3$Institute of Theoretical Physics, 
Faculty of Physics, University of
Warsaw, ul. Pasteura 5, Warsaw, Poland\\
$^4$Department of Physics, Faculty of Science, Tokyo University of Science, 
1-3, Kagurazaka,
Shinjuku-ku, Tokyo 162-8601, Japan}

\date{\today}

\begin{abstract}

We study the evolution of Gravitational Waves (GWs) during and after inflation 
as well as the resulting observational consequences in a Lorentz-violating massive 
gravity theory with one scalar (inflaton) and two tensor degrees of freedom. 
We consider two explicit examples of the tensor mass $m_g$ that depends  
either on the inflaton field $\phi$ or on its time derivative $\dot{\phi}$, 
both of which lead to parametric excitations of GWs during reheating after inflation.
The first example is Starobinsky's $R^2$ inflation model with a
 $\phi$-dependent $m_g$  and the second is a low-energy-scale inflation
model with a $\dot{\phi}$-dependent $m_g$.
We compute the energy density spectrum $\Omega_{\rm GW}(k)$ today of the GW background. 
In the Starobinsky's model, 
we show that the GWs can be amplified up to the detectable ranges of both CMB and DECIGO, but the bound from  
the big bang nucleosynthesis is quite tight to limit the growth.
In low-scale inflation with a fast transition to the reheating stage driven by the potential 
$V(\phi)=M^2 \phi^2/2$ around $\phi \approx M_{\rm pl}$ 
(where $M_{\rm pl}$ is the reduced Planck mass), 
we find that the peak position of $\Omega_{\rm GW}(k)$ induced by
 the parametric resonance can reach the sensitivity region of 
advanced LIGO for the Hubble parameter of order 1 GeV at the end of inflation.
Thus, our massive gravity scenario offers exciting possibilities for probing 
the physics of primordial GWs at various different frequencies.

\end{abstract}


\maketitle

\section{Introduction}

The inflationary paradigm successfully addresses several 
problems of the big-bang cosmology \cite{Sta80,oldinf}. 
But most importantly it gives a unique mechanism to
generate primordial cosmological perturbations that
give rise to the structure of the Universe~\cite{MukChib,oldper}. 
In the standard picture, the cosmic acceleration is driven by 
the potential energy of a scalar degree of freedom $\phi$ 
(dubbed ``inflaton''). 
The quantum fluctuations of the inflaton, which are stretched over 
the scales greater than the Hubble radius, are the source of 
the primordial scalar-type curvature perturbation \cite{MukChib,KS}.
The amplitude and the spectral index of the scalar perturbation 
are now tightly constrained by the Planck CMB 
measurements~\cite{Planckinf}.

Besides the scalar perturbation, the tensor perturbation is
also generated during 
inflation \cite{GWsta}\footnote{Throughout this paper,  by ``scalar, vector, 
and tensor" we mean those with respect to the symmetry of 3-space.}. 
In the single-field scenario with the massless tensor mode,
the tensor-to-scalar ratio $r={\cal P}_T(k)/{\cal P}_{\cal R}(k)$ is 
related to the slow-roll 
parameter $\epsilon~(=-\dot H/H^2)$, as $r=16\epsilon$ \cite{review}. 
Since we require the condition $\epsilon \ll 1$ for realizing 
inflation, the amplitude of the tensor perturbation 
is smaller than that of the scalar perturbation. 
The CMB observations so far have placed only 
the upper bound of the tensor-to-scalar 
ratio, as $r<0.11$ (95 \%\,CL) \cite{Planckinf}. 

Depending on models of inflation, the values of $\epsilon$ 
associated with the observed CMB scale are different. 
In the so-called small-field inflation where the variation 
of $\phi$ 
during inflation does not exceed the reduced Planck mass 
$M_{\rm pl}=2.435\times 10^{18}$~GeV, we have
$\epsilon\lesssim10^{-4}$ \cite{Lyth,Bau,Tsuji14}.
In Starobinsky's model \cite{Sta80}, $\epsilon\sim10^{-4}$ \cite{DT10}, 
so it marginally belongs to small-field inflation. 
The next target of CMB B-mode
measurements like LiteBIRD~\cite{LiteBIRD} and 
the ground-based Stage-4 effort \cite{stage4} is the detection of 
primordial GWs with $r$ down to the order of $10^{-3}$.

There are inflationary models in which $\epsilon\ll10^{-4}$, 
e.g., those arising from string theory \cite{sinflation}. 
In such models, it is usually believed that the detection of 
primordial GWs is impossible even with future high-precision 
CMB B-mode measurements.  However, this is not necessarily 
the case if the tensor perturbation is subject to 
some growth after inflation. 
Indeed, there are models in which the existence 
of a time-dependent tensor mass leads to the amplification
 of GWs during the reheating stage after 
inflation \cite{Lin15}\footnote{There are also models in which the 
anisotropic stress of scalar fields work as a source term 
for the production of GWs during reheating \cite{GWpre}. 
In this paper, we will not consider such a scenario.}. 

The massive gravity scenario studied in Ref.~\cite{Lin15}, which was 
originally advocated in Refs.~\cite{Dubovsky1,Dubovsky2}, 
is constructed under the internal $SO(3)$ symmetry $\phi^i\to\Lambda^i_{~j}\phi^j$,
 and $\phi^i\to\phi^i+\Xi^i(\phi^0)$, where $\Xi^i$ is 
a general function of its argument. 
The tensor modes acquire a mass due to the non-trivial vacuum 
expectation value of these four scalar fields $\phi^\mu=x^\mu$.
This property differs from the original Fierz-Pauli theory \cite{Fierz} 
and its nonlinear extension \cite{dRGT} in which the scalar field configuration 
respects Poincare symmetry. 
The internal symmetry $\phi^i\to\phi^i+\Xi^i(\phi^0)$ 
 forbids the propagation of vector modes \cite{Dubovsky1,Dubovsky2}, 
so we are left with only three dynamical 
degrees of freedom: one scalar mode and two tensor 
modes.\footnote{Throughout the paper, we avoid using the terminology ``graviton'' for the description of our tensor modes because the graviton, as a massive spin 2 particle, must have five propagating degrees of freedom due to
the Poincare symmetry in the Minkowskian space-time, which is broken in our model.}

We minimally extend our inflationary scenario by adding this tensor mass $m_g$ 
into our theory, and identify the scalar mode with the inflaton 
scalar field $\phi$ \cite{Lin15}. Since the scalar mode does not 
give rise to a ghost state, our theory
is free from the Higuchi bound \cite{Higuchi:1986py},
which is always an issue in cosmology for theories that 
respect de Sitter symmetry.
Further we impose the global scaling 
symmetry $\phi^i\to\lambda\phi^i$, which guarantees the non-dynamical nature
of $\phi^i$ in the cosmological background~\cite{Lin15,Domenech}.

Generally the tensor mass $m_g$ can depend on the 
field $\phi$ or its time derivative $\dot{\phi}$. 
To recover the local Lorentz-invariance after reheating, 
we require that $m_g$ eventually vanishes. 
The $\phi$-dependent tensor mass leads to the parametric resonant amplification 
of GWs at the early stage of reheating \cite{KLS94,KLS97} 
due to coherent oscillations of the inflaton around 
the potential minimum \cite{Lin15}. 

The parametric amplification in our massive gravity theory 
will give rise to several distinct observational signatures in CMB and 
direct measurements of GWs. Even for small-field inflation in which
 the slow-roll parameter $\epsilon$ is very small, $\epsilon\ll10^{-4}$, 
the GWs can be amplified to the detectable level in near-future
CMB experiments ($r\gtrsim 10^{-3}$).
Moreover, this parametric excitation is at work down to scales 
smaller than the Hubble radius at the onset of reheating. 
This may offer the possibility of detecting GWs in direct measurements 
such as Advanced-LIGO (A-LIGO) \cite{ALIGO} and DECIGO \cite{DECIGO}.

In this paper, we make a thorough, quantitative analysis of
the parametric resonance in two typical, explicit models
of the tensor mass term.
We consider small-field inflationary models
with two different forms of $m_g$ that depends on either $\phi$ or $\dot{\phi}$ 
and compute the primordial tensor power spectrum ${\cal P}_T$ after the amplification as well as the today's energy density spectrum $\Omega_{\rm GW}(k)$ 
of the GW background.
In Starobinsky's model with a $\phi$-dependent $m_g$, 
we will discuss the possibility of amplifying GWs at the 
detectable level in both CMB and DECIGO measurements.
In this model, however, the peak position of 
$\Omega_{\rm GW}(k)$ 
induced by the parametric excitation of GWs is at frequency 
much larger than the frequency bands of A-LIGO.

We will also study a low-scale inflation model with a 
$\dot{\phi}$-dependent $m_g$.
We consider an inflationary scenario where a fast transition 
from a nearly flat potential to the reheating stage driven by the potential $V(\phi)=M^2 \phi^2/2$ occurs around $\phi$ of the 
order of $M_{\rm pl}$. 
For the potentials where the field value
$\phi_{\rm end}$ at the end of inflation is much smaller 
than $M_{\rm pl}$ the parametric resonance is less efficient, so we focus on the models with 
$\phi_{\rm end}={\cal O}(M_{\rm pl})$.  
We will show that, as the Hubble expansion rate $H_i$ at the onset of 
reheating gets smaller, the peak position of $\Omega_{\rm GW}(k)$ shifts 
toward smaller frequencies.
For $H_i \lesssim 1$~GeV, the peak may reach 
the sensitivity band of A-LIGO.

This paper is organized as follows.
In Sec.~\ref{modelsec}, we briefly review the Lorentz-violating massive 
gravity model and present explicit forms of the tensor mass that
lead to the amplification of GWs during reheating.
In Sec.~\ref{prisec}, we derive the primordial tensor power 
spectrum generated right after the end of inflation.
In Sec.~\ref{parasec}, we discuss how the parametric resonance of 
GWs occurs during reheating for two different tensor masses.
In Secs.~\ref{Stasec} and \ref{lowsec} we numerically calculate the GW 
power spectra after the amplification in Starobinsky's model with 
a $\phi$-dependent $m_g$ and in a low-scale inflation model with 
a $\dot{\phi}$-dependent $m_g$, respectively.
In Sec.~\ref{GWbackground}, we compute the 
today's power spectrum 
of the GW background in the two models and discuss the possibility 
of detecting the GWs in direct measurements.
Sec.~\ref{consec} is devoted to conclusions.

\section{massive tensor gravity and inflation}
\label{modelsec}

The Lorentz-violating massive gravity theory proposed in 
Refs.~\cite{Dubovsky1,Dubovsky2} contains four scalar 
Goldstone fields $\phi^0$ and $\phi^i$ respecting 
the internal symmetry 
\beq\label{internalsym}
\phi^i\to\Lambda^i_{~j}\phi^j,\qquad\phi^i\to\phi^i+\Xi^i(\phi^0)\,,
\eeq
where $\Xi^i$ is a general function of its argument.
The tensor mass arises from the scalar fields' 
non-trivial vacuum expectation values (VEVs)
\be
\phi^0=t\,,\qquad 
\phi^i=x^i\,.
\label{phi0i}
\ee
At the level of lowest dimensional operators, 
there are two ingredients respecting the internal symmetry \cite{Dubovsky1},
\beq
\hspace{-0.5cm}
X &=&
g^{\mu \nu} \partial_{\mu} \phi^0 \partial_{\nu} \phi^0\,,\\
\hspace{-0.5cm}
Z^{ij} &=& 
g^{\mu \nu} \partial_{\mu} \phi^i \partial_{\nu} \phi^j
-\frac{g^{\mu \nu} \partial_{\mu} \phi^0 
\partial_{\nu} \phi^i g^{\lambda \rho} 
\partial_{\lambda} \phi^0 \partial_{\rho} \phi^j}{X}\,, 
\eeq
where $g^{\mu \nu}$ is the metric tensor. 
In the gauge where the scalar field values are fixed at its VEVs,
$\phi^\mu=x^\mu$, these quantities reduce to 
$X=-N^{-2}$ and 
$Z^{ij}=h^{ij}$, respectively, where $h^{ij}$ is the 
three-dimensional induced metric in terms of 
the Arnowitt-Deser-Misner (ADM) decomposition. 

The above model motivates us to consider a model that minimally
modifies gravity when applied to cosmology.
We first identify $\phi^0$ with the inflaton, $\phi=\phi^0$.
Then we introduce a traceless tensor constructed from 
$Z^{ij}$, as \cite{Labun}
\be
\bar{\delta }Z^{ij} \equiv 
Z^{ij}-\frac{3\delta_{kl}Z^{ik}Z^{jl}}{Z}\,,
\ee
where $Z \equiv Z^{ij} \delta_{ij}$, and consider
the action \cite{Lin15,Domenech}
\beq
S &=& 
\int d^4x \sqrt{-g} \biggl[ \frac{M_{\rm pl}^2}{2}R
-\frac{1}{2}g^{\mu \nu}\partial_{\mu}\phi \partial_{\nu}\phi
-V(\phi) \nonumber \\
&&\qquad \qquad \quad 
\,-\frac{9}{8}M_{\rm pl}^2 m_g^2
\frac{\left(\bar{\delta }Z^{ij} \right)^2}{Z^2} \biggr]\,,
\label{action}
\eeq
where $R$ is the four-dimensional Ricci scalar, $V(\phi)$ 
is the potential of the scalar field $\phi$, $m_g$ is the 
tensor mass dynamically changing in time, 
and $( \bar{\delta }Z^{ij} )^2 \equiv 
\delta_{ik} \delta_{jl}\bar{\delta }Z^{ij}
\bar{\delta }Z^{kl}$. 
In addition to the symmetry (\ref{internalsym}), the above
action has an additional global symmetry,
\begin{align}
\phi^i\to\lambda\phi^i\,,
\label{rescale}
\end{align}
where $\lambda$ is a constant.\footnote{It may be noted
that if $m_g^2=m_g^2(X)$ and $V=0$ at tree level,
the action becomes shift symmetric,
$\phi^\mu\to\phi^\mu+$constant.} 

On the flat Friedmann-Lema\^{i}tre-Robertson-Walker (FLRW) 
background described by the line-element 
$ds^2=-dt^2+a^2(t)\delta_{ij}dx^idx^j$, 
where $a(t)$ is the time-dependent scale factor, 
the rescaling symmetry (\ref{rescale}) guarantees
that the VEVs of $\phi^i$ may be identified with the
comoving coordinates $\phi^i=x^i$, and $\phi^i$
remains non-dynamical~\cite{Lin15,Domenech}.
The dynamical equations of motion are given by 
\beq
& &
3M_{\rm pl}^2H^2=\frac12 \dot{\phi}^2+V\,,
\label{back1}\\
& &
2M_{\rm pl}^2\dot{H}=-\dot{\phi}^2\,,\\
& &
\ddot{\phi}+3H\dot{\phi}+V_{,\phi}=0\,,
\label{back3}
\eeq
where an overdot represents a derivative with respect to $t$, 
$H \equiv \dot{a}/a$ is the Hubble parameter, and 
$V_{,\phi} \equiv dV/d\phi$. Note that the tensor mass term
in the action~(\ref{action}) does not 
affect the background equations of motion.

As usual, since the inflaton evolves slowly along a nearly flat 
potential during inflation, Eqs.~(\ref{back1}) and 
(\ref{back3}) approximately reduce to  
$3M_{\rm pl}^2H^2 \simeq V$ and $3H\dot{\phi}
\simeq -V_{,\phi}$, respectively. 
We define the slow-roll parameter, 
\be
\epsilon  \equiv -\frac{\dot{H}}{H^2} 
\simeq \frac{3\dot{\phi}^2}{2V} \simeq \epsilon_V\,,
\quad \epsilon_V\equiv\frac{M_{\rm pl}^2}{2}\left( 
\frac{V_{,\phi}}{V}\right)^2\,,
\label{epdef}
\ee
which is much smaller than unity during inflation. 

We assume that the tensor mass $m_g$ depends on
either $\phi$ or $\dot{\phi}$.  
For the so-called broad parametric resonance to occur
during the stage when the inflaton undergoes damped
oscillations, we require that $m_g^2$ is larger than 
$H^2$ \cite{KLS97}.
On the other hand, if $m_g^2 \gtrsim H^2$ during inflation, 
the tensor perturbation is subject to strong suppression 
by the heavy tensor mass (see e.g., Refs.~\cite{mpreheating}). 
To avoid this suppression, we consider 
the case in which $m_g^2/H^2\ll1$ during most stage of 
inflation and $m_g^2/H^2$ quickly grows to the value larger than the order of 
unity around the end of inflation.
It is possible to realize such changes 
for the following two examples:
\beq
& &
{(\rm i)}~~m_g^2(\phi)=\lambda \phi^2 
e^{-b (\phi/M_{\rm pl})^n}\,,
\label{mg1}\\
& &
{(\rm ii)}~~m_g^2(\dot{\phi})=\mu \frac{\dot{\phi}^2}
{M_{\rm pl}^2}\,, 
\label{mg2}
\eeq
where $\lambda$, $b$, $n$, $\mu$ are positive 
dimensionless constants.
For the potential with a minimum at $\phi=0$, 
the tensor mass in the case (i) rapidly approaches 0 
as the inflaton decays to radiation. 
This property also persists in the case of (ii). 
Hence the current observational bounds of 
the tensor mass \cite{grabound} can be safely
satisfied for the above two choices of $m_g^2$.

In the case (i), if inflation occurs in the region where 
$\phi$ is larger than $M_{\rm pl}$, the exponential factor 
in Eq.~(\ref{mg1}) with $b={\cal O}(1)$ 
can suppress $m_g^2(\phi)$ so that $m_g^2(\phi)/H^2\ll1$ during inflation. 
The ratio $m_g^2(\phi)/H^2$ can grow with the decrease of 
$\phi$ toward the end of inflation, so this allows 
the possibility to realize the broad parametric resonance 
driven by the oscillating tensor mass 
squared $m_g^2 (\phi) \simeq \lambda \phi^2$ 
around $\phi=0$. 

In the case (ii), the ratio between $m_g^2 (\dot{\phi}^2)$ and 
$H^2$ during inflation can be estimated as 
\be
\frac{m_g^2 (\dot{\phi}^2)}{H^2} 
\simeq 2\mu \epsilon\,,
\label{mgratio}
\ee
which is smaller than unity for $\mu \epsilon \ll 1$.
The slow-roll parameter $\epsilon$ grows to the order 
of unity by the end of inflation, so the ratio 
$m_g^2 (\dot{\phi}^2)/H^2$ becomes larger than unity
for $\mu \gtrsim 1$ around the onset of reheating.
Thus, the broad parametric resonance can occur 
for $\mu \gg 1$.

\section{Primordial power spectra generated during inflation}
\label{prisec}

We consider a linearly perturbed line-element on 
the flat FLRW background, as 
\beq
ds^2&=&
-(1+2\alpha)dt^2+2a(t) \left( \beta_{|i}+S_i \right) 
dt dx^i \nonumber \\
& & +a^2(t) \left[ (1+2\psi)\delta_{ij}+2E_{|ij}
+2F_{i|j}+\gamma_{ij} \right]dx^idx^j, \nonumber \\
\eeq
where the subscript $|i$ represents a three-dimensional covariant
 derivative, $\alpha,\beta,\psi,E$ are scalar metric 
perturbations, $S_i,F_i$ are vector perturbations satisfying
the transverse conditions ${S_i}^{|i}=0,{F_i}^{|i}=0$, 
and $\gamma_{ij}$ is the tensor perturbation obeying the 
transverse and traceless conditions 
${\gamma_{ij}}^{|j}=0, {\gamma_{i}}^{i}=0$. 
Since the vector perturbation does not propagate in our 
massive gravity theory due to the internal 
symmetry (\ref{rescale}) \cite{Lin15,Domenech}, 
we will consider the propagation of scalar and tensor 
perturbations in the following.

\subsection{Scalar power spectrum}

At the level of linear cosmological perturbations, the tensor 
mass term on the r.h.s. of Eq.~(\ref{action}) does not 
modify the dynamics of scalar perturbations. 
We define the curvature perturbation, as 
${\cal R}=\psi-H\delta \phi/\dot{\phi}$, where 
$\delta \phi$ is the perturbation of $\phi$. 
Since $\delta \phi=0$ in the unitary gauge (\ref{phi0i}), 
${\cal R}$ is equivalent to $\psi$.
The primordial power spectrum of ${\cal R}$ generated 
right after the end of inflation is given by \cite{review}
\be
{\cal P}_{\cal R}=\frac{H^2}{8\pi^2 \epsilon 
M_{\rm pl}^2}\bigg|_{k=aH}\,,
\label{scapower}
\ee
where $k$ is the comoving wavenumber and 
$\epsilon$ is given by Eq.~(\ref{epdef}).
The power spectrum (\ref{scapower}) should be evaluated at the 
Hubble horizon crossing ($k=aH$) during inflation. 
The scalar spectral index is
\be
n_s-1 \equiv \frac{d \ln {\cal P}_{\cal R}}{d\ln k}
\biggr|_{k=aH}
=-6\epsilon_V+2\eta_V\,,
\label{scalar}
\ee
where $\epsilon_V$ is defined in Eq.~(\ref{epdef}), and 
\be
\eta_V \equiv \frac{M_{\rm pl}^2 V_{,\phi \phi}}{V}\,.
\label{eta}
\ee
{}From the CMB observations, the amplitude of the primordial
 scalar power spectrum (\ref{scapower}) is constrained to be 
${\cal P}_{\cal R} \simeq 2.2 \times 10^{-9}$ for the 
perturbation which crossed the Hubble radius about the 
number of e-folding $N=55$ before the end of inflation \cite{Planckinf}. 
This implies that the Hubble parameter during inflation 
(denoted as $H_{\rm inf}$)
is expressed in terms of the slow-roll parameter $\epsilon$ as 
\be
H_{\rm inf} \simeq \sqrt{\epsilon} \times 
10^{15}~{\rm GeV}\,.
\label{Hinf}
\ee
Thus smaller the $\epsilon$ is, lower 
the energy scale of inflation becomes. 

\subsection{Tensor perturbations}

The tensor mass term in the action~(\ref{action}) 
gives rise to modifications to the dynamics of 
the tensor perturbation.
Expanding the action (\ref{action}) up to quadratic order 
in perturbation yields the second-order action of 
$\gamma_{ij}$,
\begin{align}
S_T^{(2)}=&\frac{M_{\rm pl}^2}{8}
\int d^4x a^3 \delta^{ik} \delta^{jl}
\cr
&\times
\left({\dot\gamma}_{ij}{\dot\gamma}_{kl}
+\gamma_{ij}\nabla^2\gamma_{kl}
-m_g^2 \gamma_{ij}\gamma_{kl}\right),
\label{Taction}
\end{align}
where $\nabla^2 \equiv\delta^{ij}\partial_i \partial_j/a^2$. 
The resulting equation of motion for $\gamma_{ij}$ 
in real space reads 
\be
{\ddot\gamma}_{ij}+3H\dot{\gamma}_{ij}
-{\nabla^2}\gamma_{ij}+m_g^2\gamma_{ij}=0\,.
\ee
We decompose the field $\gamma_{ij}$ into Fourier modes as
\be
\gamma_{ij}({\bm x},t)= \int \frac{d^3k}{(2\pi)^{3/2}}
e^{i\bm{k}\cdot\bm{x}}\hat{\gamma}_{ij}({\bm k},t)\,,
\label{eq:tensormodes}
\ee
where ${\bm k}$ is a comoving wavenumber, and 
\be
\hat{\gamma}_{ij}({\bm k},t)=
\sum_{s=+,\times} 
[\gamma(k,t) a_{s} ({\bm k})+
\gamma^* (k,t) a_{s}^{\dagger} (-{\bm k})] \epsilon_{ij}^{(s)} ({\bm k}),
\label{gammaij}
\ee
with $s=+,\times$ being the two polarization states.
The polarization tensors $\epsilon_{ij}^{(s)}({\bm k})$, which  
are transverse and traceless 
($k^j \epsilon_{ij}^{(s)}=\delta^{ij}\epsilon_{ij}^{(s)}=0$), 
satisfy the normalization $\delta^{ik}\delta^{jl}\epsilon^{(s)}_{ij} (\bm{k})
\epsilon^{*(s')}_{kl} (\bm{k})=\delta_{s s'}$.
The annihilation and creation operators
$a_{s} ({\bm k})$ and $a_{s}^{\dagger} ({\bm k}')$ obey
 the commutation relation 
$[a_{s}({\bm k}),a_{s'}^{\dagger}({\bm k}')]=\delta_{ss'}\delta^{(3)}({\bm k}-{\bm k}')$.
The primordial tensor power spectrum ${\cal P}_T(k,t)$ 
per unit logarithmic frequency interval is given by
\be
{\cal P}_T (k,t)=2\cdot \frac{k^3}{2\pi^2} |\gamma(k,t)|^2\,,
\label{powerspe}
\ee
where the factor 2 in front comes from the two 
polarization states.

We introduce a canonically normalized field,
\be
\v_{ij}\equiv \frac{M_{\rm pl}}{2} \gamma_{ij}\,,
\label{vfield}
\ee
and the corresponding mode functions in the 
momentum space,
\be
\v(k,t)\equiv \frac{M_{\rm pl}}{2} \gamma(k,t)\,.
\label{vsdef}
\ee
Then $v$ obeys the equation of motion,
\begin{align}
\ddot{v}+3H\dot v+\left(\frac{k^2}{a^2}+m_g^2\right)v=0\,,
\label{vseqt}
\end{align}
or in terms of the conformal time $\tau=\int^t \,dt/a$,
\be
v''+2\calH v'+\left( k^2+m_g^2a^2 \right)v=0\,,
\label{vseq}
\ee
where a prime represents the derivative with respect to 
conformal time, ${}'=\partial/\partial\tau$ 
and ${\cal H}=a'/a$.
It satisfies the Klein-Gordon normalization,
\be
v\,v^*{}'-v'v^*=\frac{i}{a^2}\,.
\label{KGnorm}
\ee
 The mode function that satisfies the above is called
a positive frequency function.

\subsection{Tensor power spectrum for $m_g^2/H^2\ll 1$}

Assuming that $m_g^2\ll H^2$ and $|\dot{m}_g/(Hm_g)| \ll1$
during inflation and adopting the de Sitter background approximation,
$a\simeq (-H\tau)^{-1}$,
which is valid as long as the wavelength is shorter
than the Hubble radius, $k/a\gtrsim H$ ($\leftrightarrow k\gtrsim\calH$),
the natural positive frequency function in the limit
$k\gg\calH$ is given by
\be
v=\frac{1}{\sqrt{2(k/a)a^3}}e^{-i\int^{t}dt\,k/a}
=\frac{1}{a\sqrt{2k}}e^{-ik\tau}\,.
\label{WKBsol}
\ee
This solution is valid as long as 
$k/\calH\approx(-k\tau)\gtrsim1$,
and coincides with the mode function of 
the Bunch-Davies vacuum 
in the pure de Sitter limit.

As the Universe expands and the comoving 
wavelength becomes greater than the Hubble radius, 
$k/\calH\approx(-k\tau)\lesssim 1$,
the de Sitter approximation is no longer valid. 
In this case, on the other hand, the perturbation $v$ 
ceases its oscillations and the solution is given by the friction-dominated slow-roll solution,
\be
v \simeq C\exp\left[-\int^t\frac{m_g^2}{3H}dt\right]\,.
\label{SHsol}
\ee
The constant $C$ is determined by matching the two solutions (\ref{WKBsol}) and (\ref{SHsol}) at 
the horizon crossing time $t_k$ characterized by 
$k=a(t_k)H(t_k)$
($\leftrightarrow k/\calH(\tau_k)\approx -k\tau_{k}=1$).
This leads to the solution for $t>t_k$, 
\be
v(k,t)=\frac{H(t_k)}{\sqrt{2k^3}}
\exp\left[-\int_{t_k}^t\frac{m_g^2}{3H}dt\right]\,,
\label{SHmatched}
\ee
apart from an irrelevant phase factor.

The tensor power spectrum (\ref{powerspe}) is then given by
\be
{\cal P}_T(k,t)=
\frac{2k^2}{\pi^2 M_{\rm pl}^2 a^2}\,,
\quad\text{for}~k \gg aH\,,
\label{Phge1}
\ee
and
\be
{\cal P}_T(k,t)=\frac{2H^2(t_k)}{\pi^2M_{\rm pl}^2} 
\exp\left[-\int_{t_k}^t\frac{2m_g^2}{3H}dt\right]\,,
\quad\text{for}~k \ll aH\,.
\label{Phge2}
\ee

If $m_g^2/H^2$ remains small until
the end of inflation, say until $t=t_f$,
and its effect could be neglected after inflation, 
the spectrum for the modes $k\ll aH$ becomes
\begin{align}
{\cal P}_T(k,t_f)=\frac{2H^2(N_k)}{\pi^2M_{\rm pl}^2} 
\exp\left[-\int_{0}^{N_k}\frac{2m_g^2}{3H^2}dN\right]\,,
\label{Phs}
\end{align}
where $dN=-Hdt$, and $N_k$ is the number of e-folding
at Hubble horizon crossing counted backward from the end of
inflation, $N_k=\int_{t_k}^{t_f}Hdt$.
The spectrum index is given by
\be
n_T\equiv\frac{d\ln{\cal P}_T}{d\ln k}
=-2\epsilon(t_k)+\frac{2m_g^2}{3H^2}(t_k)\,.
\label{nts}
\ee
The tensor-to-scalar ratio reads
\be
r=16\epsilon(t_k)
\exp\left[-\int_{0}^{N_k}\frac{2m_g^2}{3H^2}dN\right]\,.
\label{rs}
\ee
Thus the blue-tilted spectrum may be realized if $m_g^2/H^2>3\epsilon$
at the time of Hubble horizon crossing. However, due to the
evolution of the massive tensor on super-horizon scales,
the tensor-to-scalar ratio at CMB observation scales,
$N_k\gg1$, would be substantially suppressed in comparison with 
the massless tensor case.

\subsection{Tensor spectrum after 
the transition from $m_g^2\ll H^2$ to $m_g^2\gg H^2$ 
during inflation} 
\label{trasec}

If the tensor mass is too heavy ($m_g^2/H^2\gg1$) during inflation, 
the tensor perturbation is subject to strong suppression. 
To avoid such suppression, we need $m_g^2/H^2\ll1$ 
during the most stage of inflation. 
For the broad parametric resonance to occur right after inflation, 
on the other hand, the ratio $m_g^2/H^2$ must be larger than unity at the onset of reheating.
Thus, we consider the case where the transition from the almost
massless regime $m_g^2/H^2\ll1$ to the massive regime $m_g^2/H^2\gg1$ occurs at the late stage of inflation, and solve the evolution of the mode functions until the onset of inflaton oscillations.

\begin{figure}
\begin{center}
\includegraphics[width=3.5in]{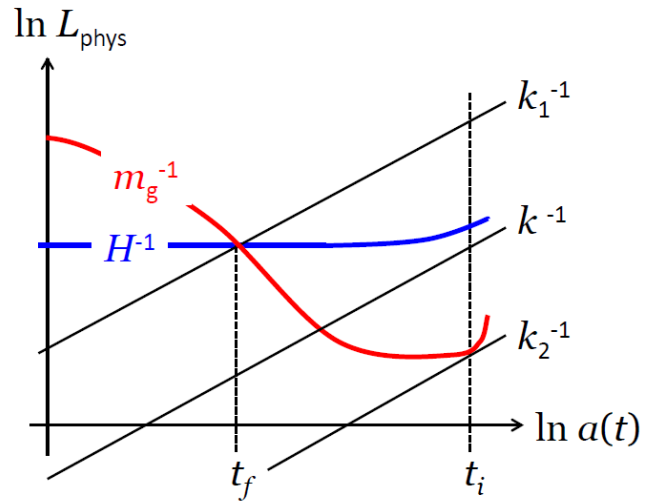}
\end{center}
\caption{\label{fig1}
The physical scales of our interest at which the
tensor power spectrum changes its shape are depicted
in the spacetime diagram. At $t>t_f$ the amplitude of
the tensor modes $v(k,t)$ outside the Hubble 
horizon is nearly constant in time, 
while the modes inside the Hubble horizon 
behave as $\propto a^{-1}$.
At $t_f<t<t_i$, the modes at $a/k>m_g^{-1}$ evolve 
as $a^{-3/2}$, while those at $a/k<m_g^{-1}$ behave 
as $a^{-1}$.}
\end{figure}

In the following, we assume that the tensor mass satisfies 
$m_g^2\ll H^2$ initially, gradually grows and crosses $m_g^2\simeq H^2$ at $t=t_f$,
and becomes $m_g^2\gg H^2$ until $t=t_i$ ($>t_f$), 
where $t_i$ is the time at the onset of the 
oscillatory stage of $\phi$.
For the sake of analytical estimates, we employ the approximation
that $m_g^2\ll H^2$ at $t<t_f$, $m_g^2=H^2$ at $t=t_f$,
and $m_g^2\gg H^2$ at $t_f<t<t_i$,
which is qualitatively valid for most of the modes of interest.
For reference, the physical scales of our interest are depicted 
in Fig.~\ref{fig1}.

First, let us consider the modes 
$k<m_g(t_f) a(t_f)=H(t_f) a(t_f)\equiv k_1$.
For these modes, since $k/a\ll H\ll m_g$ for $t>t_f$,
Eq.~(\ref{vseqt}) may be easily solved to give
\be
v(k,t)=
v(k,t_f)\left[\frac{a(t_f)}{a(t)}\right]^{3/2}
\left[\frac{m_g(t_f)}{m_g(t)}\right]^{1/2}
e^{\pm i\int^t m_g dt},
\label{vsolk1}
\ee
where $e^{\pm i\int^t m_g dt}$ is an abbreviation
for $\alpha e^{-i\int^t m_g dt}+\beta e^{+i\int^t m_g dt}$
with $|\alpha|^2+|\beta|^2=1$. We adopt the above
notation since the explicit values of $\alpha$ and $\beta$
are unnecessary for our discussion.
The value of $v(k,t_f)$ is given by Eq.~(\ref{SHmatched}) 
with $t=t_f$. Taking into account the variation of $H$ from 
$t=t_k$ ($\leftrightarrow N=N_k$) to $t=t_f$ 
($\leftrightarrow N=N_f$), it follows that 
$H(t_k)=H(t_f)\exp(\int_{N_f}^{N_k} \epsilon\,dN)$. 
Then, we obtain 
\be
v(k,t_f)=\frac{H(t_f)}{\sqrt{2k^3}}
\exp\left[\int_{N_f}^{N_k}\left(\epsilon-\frac{m_g^2}{3H^2}\right)dN\right]\,.
\label{k0k1}
\ee
The resulting power spectrum at the end of 
inflation  is given by 
\beq
& &
{\cal P}_T(k,t_i)
={\cal P}_T(k_1,t_i)
\exp\left[ 2\int_{N_f}^{N_k}\left(\epsilon
-\frac{m_g^2}{3H^2}\right)
dN\right]\,, \nonumber \\
& &
\qquad \qquad \quad~
{\rm for}~k<k_1\,,
\label{Phma1}
\eeq
where
\begin{align}
{\cal P}_T(k_1,t_i)
=\frac{2H^2(t_f)}{\pi^2M_{\rm pl}^2}
\frac{m_g(t_f)}{m_g(t_i)}e^{-3N_{f \to i}}\,,
\end{align}
and $N_{f\to i}\equiv\ln [a(t_i)/a(t_f)\bigr]$.
Note that $H(t_f)=m_g(t_f)$ by definition.

Let us now turn to the modes in the range
$k_1<k<k_2$ where $k_2\equiv m_g(t_i)a(t_i)$.  
The highest wavenumber
is set at $k=k_2$ because the effect of the mass term would be 
negligible for the modes with $k>k_2$ at any stage of
the Universe. For the modes $k_1<k<k_2$ the frequency 
$\omega_k \equiv \sqrt{k^2/a^2+m_g^2}$ is much larger 
than $H$ by the end of inflation, so the solution to Eq.~(\ref{vseqt}) yields
\be
v(k,t)=\frac{1}{a^{3/2} \sqrt{2\omega_k}} 
e^{\pm i\int^t \omega_k dt}\,.
\label{vkin}
\ee
The perturbation in the range $k_1<k<k_2$ crosses $k=m_g a$ at an epoch,
say $t=t_m$, where $t_m$ is in the range $t_f<t_m<t_i$.
For $t<t_m$ we have $k^2/a^2 \gg m_g^2$, so Eq.~(\ref{vkin}) 
reproduces the early-time vacuum solution (\ref{WKBsol}) with  
$|v(k,t)| \propto a^{-1}$.
For $t>t_m$ we have $\omega_k \simeq m_g$ and 
hence $|v(k,t)| \propto a^{-3/2}$ from Eq.~(\ref{vkin}).
The resulting amplitude at the end of inflation is given by 
\be
\left| v(k,t_i) \right|=\frac{1}{a(t_i)^{3/2}\sqrt{2m_g(t_i)}}\,,
\ee
which leads to the highly blue-tilted spectrum 
\be
{\cal P}_T(k,t_i)={\cal P}_T(k_1,t_i)
\left( \frac{k}{k_1} \right)^3\,,
\quad {\rm for}~ k_1<k<k_2\,.
\label{Phma2}
\ee

For the modes $k>k_2$ the solution to Eq.~(\ref{vseqt}) 
is given by Eq.~(\ref{vkin}) with $\omega_k \simeq k/a$, 
so that the perturbation evolves as
$|v(k, t)| \propto a^{-1}$. The power spectrum at the end of 
inflation has the scale-dependence  
\beq
\hspace{-0.3cm}
{\cal P}_T(k,t_i)
&=&
\frac{2}{\pi^2 M_{\rm pl}^2} 
\left( \frac{k}{a(t_i)} \right)^2 \nonumber \\
\hspace{-0.3cm}
&=&{\cal P}_T(k_2,t_i)\left(\frac{k}{k_2}\right)^2\,,
\quad {\rm for}~~k>k_2\,,
\label{Phma3}
\eeq
which is smoothly matched with Eq.~(\ref{Phma2}) at 
$k=k_2=m_g(t_i)a(t_i)$, where
\begin{align}
{\cal P}_T(k_2,t_i)={\cal P}_T(k_1,t_i)\left(\frac{k_2}{k_1}\right)^3
=\frac{2m_g^2(t_i)}{\pi^2 M_{\rm pl}^2}\,.
\end{align}

In summary, the power spectrum at the end of inflation 
is given by Eq.~(\ref{Phma1}) for $k<k_1$, 
Eq.~(\ref{Phma2}) for $k_1<k<k_2$, and 
Eq.~(\ref{Phma3}) for $k>k_2$, 
As clear from the above discussion, the spectrum suffers from
strong suppression if the transition from
the almost massless stage to the massive stage takes
too many e-foldings.
In Sec.\,\ref{Stasec}, we will confirm the above analytic estimation 
in the Starobinsky model
by using the tensor mass squared (\ref{mg1}).

\section{Parametric resonance of gravitational waves 
during reheating}
\label{parasec}
 
The inflationary epoch is followed by the reheating stage in 
which the inflaton field oscillates around its potential minimum. 
If the potential has a minimum at 
$\phi=0$, it can be expanded in the following form 
\be
V(\phi)=\frac{1}{2}M^2\phi^2+\cdots\,,
\label{poexpan}
\ee
where $M$ is the inflaton mass during reheating 
and the dots stand for corrections to the leading-order 
term $M^2 \phi^2/2$. 

The end of inflation depends on the form of the potential $V(\phi)$ in the preceding inflationary epoch. In chaotic inflation where the potential is exactly given by $V(\phi)=M^2 \phi^2/2$ \cite{chaotic}, the field value at the end of inflation can be estimated as $\phi_{\rm end}=\sqrt{2}M_{\rm pl}$ from the 
condition $\epsilon_V=1$. In this case, the mass scale $M$ is of the same order as the Hubble parameter $H_i$ at the onset 
of reheating.

If we consider the potential $V(\phi)=V_0 
(1-e^{-\alpha \phi/M_{\rm pl}})^2$ of the so-called 
$\alpha$-attractor model \cite{alpha}, the leading-order 
contribution to $V(\phi)$ around $\phi=0$ is given by 
Eq.~(\ref{poexpan}) with 
$M^2=2V_0 \alpha^2/M_{\rm pl}^2$. 
The slow-roll parameter for this potential is 
$\epsilon_V=2\alpha^2 (e^{\alpha \phi/M_{\rm pl}}-1)^{-2}$, 
so $\phi_{\rm end}$ tends to be 
smaller for larger $\alpha$. 
The Starobinsky model discussed later in Sec.~\ref{Stasec} 
corresponds to $\alpha=\sqrt{6}/3$, in which case 
$\phi_{\rm end}=0.94M_{\rm pl}$. 
If $\alpha=100$, then $\phi_{\rm end}=0.05M_{\rm pl}$. 
For $\phi_{\rm end} \ll M_{\rm pl}$, it follows that 
$H_i \ll M$.
As we will see below, for smaller $\phi_{\rm end}$, the broad parametric resonance 
after inflation tends to be less efficient.

During reheating, the inflaton energy density finally 
decays to the radiation energy density $\rho_r$. 
We consider the Born decay with the friction term 
$\Gamma \dot{\phi}$ in the inflaton equation of motion, 
where $\Gamma$ is a decay constant \cite{oldre,Dolgov}. 
On the FLRW background the dynamical equations 
of motion are then given by 
\beq
& & 
3M_{\rm pl}^2H^2=
\frac{1}{2}\dot{\phi}^2+V(\phi)+\rho_r\,,
\label{Heq}\\
& & 
\ddot{\phi}+\left( 3H+\Gamma \right) \dot{\phi}+V_{,\phi}=0\,,\label{phieq}\\
& &
\dot{\rho}_r+4H\rho_r=\Gamma \dot{\phi}^2\,.
\label{rhoreq}
\eeq
The period driven by oscillations of the massive inflaton 
is characterized by the transient matter era 
in which  the scale factor evolves as $a \propto t^{2/3}$. 
At the early stage of reheating, where the condition 
$H \gg \Gamma$ is satisfied, the radiation is not yet 
sufficiently generated, so Eqs.~(\ref{Heq}) and 
(\ref{phieq}) reduce, respectively, to  
$4M_{\rm pl}^2/(3t^2) \simeq \dot{\phi}^2/2+M^2 \phi^2/2$ 
and $\ddot{\phi}+(2/t)\dot{\phi}+M^2 \phi \simeq 0$. 
The solution to $\phi$ compatible with these equations 
is given by 
\be
\phi(t) \simeq \sqrt{\frac{8}{3}} 
\frac{M_{\rm pl}}{Mt} \sin (Mt+\theta_0)\,,
\label{phiosc}
\ee
where $\theta_0$ is an arbitrary constant.  
For the inflationary models in which $\phi_{\rm end}$ is 
of the order $M_{\rm pl}$ (like chaotic inflation and 
Starobinsky model), the initial time $t_i$ 
at the onset of reheating corresponds to $t_i={\cal O}(1/M)$.
For the models with $\phi_{\rm end} \ll M_{\rm pl}$ 
(like the $\alpha$-attractor with $\alpha \gg 1$), we 
have that $Mt_i \gg 1$.

After the Hubble expansion rate $H$ drops below 
$\Gamma$, the solution to Eq.~(\ref{phieq}) changes to 
\be
\phi(t) \propto e^{-\Gamma (t-t_i)/2} 
\sin (Mt+\theta_0)\,,
\ee
where we used the condition $\Gamma \ll M$. 
Hence the field $\phi$ starts to decay rapidly to 
generate the radiation around the time defined by 
$t_\Gamma \equiv 1/\Gamma$. 
The reheating temperature is estimated as \cite{KLS97,Kuro14}
\be
T_\Gamma =1.1 g_*^{-1/4}\sqrt{\Gamma M_{\rm pl}}\,,
\label{Tg}
\ee
where $g_*$ is the relativistic degrees of freedom 
at $t_\Gamma$. 

For concreteness, let us first consider the 
$\phi$-dependent tensor mass squared given by Eq.~(\ref{mg1}) 
with $b={\cal O}(1)$ and $n={\cal O}(1)$.
For the models with $\phi_{\rm end} \lesssim M_{\rm pl}$ 
we have $b(\phi_{\rm end}/M_{\rm pl})^n \lesssim 1$, so
$m_g^2(\phi)$ can be approximated as 
$\lambda \phi^2$. Then, the coherent oscillation 
of inflaton leads to the excitation of the perturbation $v$
by the parametric resonance. 
Introducing the rescaled 
field $X_T=a^{3/2}v$ in Eq.~(\ref{vseqt}), it follows that 
\be
\ddot{X}_T+\left[ \frac{k^2}{a^2}
+\lambda \phi^2 e^{-b(\phi/M_{\rm pl})^n}
-\frac{9}{4}H^2-\frac{3}{2}\dot{H} \right]X_T=0\,.
\label{Xs}
\ee
On using the solution (\ref{phiosc}) with $\theta_0=0$, 
we can express Eq.~(\ref{Xs}) in form of the Mathieu equation 
\be
\frac{d^2 X_T}{dz^2}+\left[ A_k-2q \cos(2z) 
\right]X_T=0\,,
\label{Xs2}
\ee
where 
\beq
A_k &=& \frac{k^2}{M^2a^2}+2q
-\frac{9H^2}{4M^2}-\frac{3\dot{H}}{2M^2}\,,
\label{Ak}\\
q &=& \frac{2\lambda}{3} \left( \frac{M_{\rm pl}}
{M} \right)^2 \frac{e^{-b(\phi/M_{\rm pl})^n}}{z^2}\,,
\label{qk} \\
z &=& Mt\,.
\label{zk}
\eeq

The broad parametric resonance occurs in the 
regime where the parameter $q$ is much larger 
than 1 \cite{KLS94,KLS97}. 
Since the Hubble parameter during reheating can be estimated 
as $H^2 \simeq M^2 \phi^2/(3M_{\rm pl}^2)$, it follows that $m_g^2/H^2 \simeq 3\lambda(M_{\rm pl}/M)^2e^{-b(\phi/M_{\rm pl})^n}$.
For the models in which $\phi_{\rm end}$ is of the order 
$M_{\rm pl}$, we have $z_i=Mt_i={\cal O}(1)$ and 
hence the resonance parameter $q$ is of the similar order 
to $m_g^2/H^2$ at the onset of reheating ($t=t_i$).
In such cases, as long as the condition
\be
\lambda  \left( \frac{M_{\rm pl}}{M} \right)^2 
\gg 1
\label{lamcon}
\ee
is satisfied, both $q$ and $m_g^2/H^2$ are much larger than 1 
at $t=t_i$.
Since $q$ decreases in proportion to $t^{-2}$, the 
perturbation $X_T$ 
crosses many instability and stability bands present 
for the system of the Mathieu equation (\ref{Xs2}) \cite{KLS97,Tsuji99}. 
Even if the resonance occurs stochastically in the expanding 
Universe, the rapid movement of inflaton around $\phi=0$ 
leads to the non-adiabatic growth of tensor perturbations. 

For the models with $\phi_{\rm end} \ll M_{\rm pl}$
we have $z_i=Mt_i \gg 1$, so the parameter 
$q$ is much smaller than the ratio $m_g^2/H^2$ 
at $t=t_i$. This means that, even if $m_g^2$ grows 
to a value much larger than $H^2$ at the end of inflation, 
the parametric resonance tends to be less efficient relative 
to the case $\phi_{\rm end}={\cal O}(M_{\rm pl})$. 
In such cases, we need to choose a larger coupling 
constant $\lambda$ for the realization of the excitation 
of GWs similar to that for $\phi_{\rm end}={\cal O}(M_{\rm pl})$.

During the coherent oscillation of inflaton the term 
$k^2/(M^2 a^2)$ is in proportion to $t^{-4/3}$, 
which decreases more slowly relative to the term $2q$. 
If the condition $k^2/(M^2a_i^2)>2q_i$ is 
satisfied at the beginning of reheating 
(where the subscript ``$i$'' represents quantities at $t=t_i$), 
the gradient term $k^2/(M^2 a^2)$ dominates over the 
other terms in the square bracket of Eq.~(\ref{Xs2}) and 
hence the broad parametric resonance does not occur.
Then, we obtain the cut-off wavenumber 
\be
\frac{k_{\rm cut}}{a_iH_i}
=\sqrt{\frac{4\lambda}{3}} \frac{M_{\rm pl}}{H_i}
\frac{e^{-b(\phi_i/M_{\rm pl})^n/2}}{z_i}\,.
\label{kcut}
\ee
The GWs with wavenumbers in the range 
\be
k \lesssim k_{\rm cut}\,,
\label{kcutcon}
\ee
are subject to the parametric amplification during reheating.
Since $e^{-b(\phi_i/M_{\rm pl})^n/2}$ is of the 
order unity for $\phi_i \lesssim M_{\rm pl}$, 
we have $k_{\rm cut}/(a_iH_i) 
\approx \sqrt{\lambda}(M_{\rm pl}/H_i)z_i^{-1}$. 

The broad parametric resonance ends after $q$ drops below 
the order of 1. The narrow parametric resonance occurs in 
the instability bands ranging in the region
$0.3 \lesssim q \lesssim 0.8$ \cite{KLS97}.
The end of amplification (labelled by the subscript ``$e$'') 
is characterized by the condition $q_e \simeq 0.3$, such that  
\be
z_e \simeq 1.5 \sqrt{\lambda} \frac{M_{\rm pl}}{M}\,.
\label{zf}
\ee
The typical wavenumber $k_*$ associated with the parametric 
excitation of GWs corresponds to the mode 
$k_*^2/(M^2a_e^2) \simeq 2q_e \simeq 0.6$, i.e., 
\be
\frac{k_*}{a_iH_i} \simeq 0.8 \frac{M}{H_i} 
\left( \frac{z_e}{z_i} \right)^{3/2}\,.
\label{kstar}
\ee
For the modes $k>k_*$, the gradient term $k^2/(M^2a^2)$ starts to 
dominate over the resonance term $2q \cos(2z)$ 
before the end of amplification, so the parametric resonance 
tends to be less efficient. 
On the other hand, the resonance does not occur 
for the modes $k>k_{\rm cut}$.
If $\lambda \simeq 10^{-6}$, $M \simeq 10^{-5}M_{\rm pl}$, $H_i \simeq 0.5M$, 
and $z_i=1$, for example, we have
$k_*/(a_iH_i) \simeq 45$ and 
$k_{\rm cut}/(a_iH_i) \simeq 200$ with $z_e \simeq 150$.

We also consider the $\dot{\phi}$-dependent tensor mass squared 
given by Eq.~(\ref{mg2}). 
Employing the background solution 
(\ref{phiosc}) with $\theta_0=\pi/2$, the rescaled field 
$X_T=a^{3/2}v$ obeys the Mathieu 
equation (\ref{Xs2}) with 
\be
q=\frac{2\mu}{3z^2}\,,
\label{qmu}
\ee
where $A_k$ and $z$ are defined in the same as 
Eqs.~(\ref{Ak}) and (\ref{zk}), respectively. 
Compared to Eq.~(\ref{qk}), there is the correspondence 
$\mu  \to \lambda(M_{\rm pl}/M)^2$ and
$e^{-b(\phi/M_{\rm pl})^n} \to 1$. 
For the models with $\phi_{\rm end}=
{\cal O}(M_{\rm pl})$, i.e., $z_i=Mt_i={\cal O}(1)$, 
the broad parametric resonance occurs for 
\be
\mu \gg 1\,.
\ee
For the models with $\phi_{\rm end} \ll {\cal O}
(M_{\rm pl})$ we have $z_i \gg {\cal O}(1)$, 
so the broad resonance is less efficient 
relative to the case $\phi_{\rm end}=
{\cal O}(M_{\rm pl})$ for the same coupling $\mu$.

The discussion given between Eq.~(\ref{kcut}) and 
(\ref{kstar}) is valid by replacing $\lambda$ with 
$\mu(M/M_{\rm pl})^2$, e.g., 
$k_{\rm cut}/(a_iH_i)=\sqrt{4\mu/3}(M/H_i)z_i^{-1}$ and 
$z_e \simeq 1.5 \sqrt{\mu}$.

\section{Starobinsky inflation with the 
$\phi$-dependent tensor mass}
\label{Stasec}

For the tensor mass squared given by Eq.~(\ref{mg1}), 
we numerically compute the primordial tensor power spectrum 
at the end of reheating for the inflaton potential 
\be
V(\phi)=\frac34 M^2M_{\rm pl}^2 \left[ 
1- e^{-\sqrt{6}\phi/(3M_{\rm pl})}\right]^2\,.
\label{poten}
\ee
This follows from the Lagrangian 
$f(R)=R+R^2/(6M^2)$ after a conformal transformation 
to the Einstein frame \cite{DT10}. 
The scalar degree of freedom $\phi$ 
is related to the Ricci scalar $R$, as 
$\phi=\sqrt{3/2}\,M_{\rm pl} \ln [1+R/(3M^2)]$.
Expanding the potential (\ref{poten}) around $\phi=0$, 
the leading-order contribution corresponds 
to $M^2 \phi^2/2$.

For the potential (\ref{poten}), the slow-roll parameter (\ref{epdef}) is given by 
\be
\epsilon_V = \frac43 \left[  
e^{\sqrt{6} \phi/(3M_{\rm pl})}-1 
\right]^{-2}\,,
\label{ep2}
\ee
so that the end of inflation ($\epsilon_V=1$) corresponds to 
the field value $\phi_{\rm end}=0.94M_{\rm pl}$. 
The inflationary expansion is realized in the regime $\phi \gtrsim M_{\rm pl}$.
The e-folding number associated with the field 
value $\phi$ during inflation can be estimated as 
\be
N=\frac{1}{M_{\rm pl}^2}
\int_{\phi_{\rm end}}^{\phi} 
\frac{V}{V_{,\tilde{\phi}}} d\tilde{\phi} 
\simeq \frac34
e^{\sqrt{6}\phi/(3M_{\rm pl})}
-\frac{\sqrt{6}\phi}{4 M_{\rm pl}}\,,
\label{efold}
\ee
where we neglected the contribution 
arising from $\phi_{\rm end}$. 
Taking the dominant contributions in Eqs.~(\ref{ep2}) 
and (\ref{efold}), we obtain the following relation 
\be
\epsilon \simeq \frac{3}{4 N^2}\,.
\ee
For $N=55$, we have $\epsilon \simeq 2.5 \times 10^{-4}$, so 
the Hubble parameter in Starobinsky inflation corresponds to 
$H_{\rm inf} \simeq 10^{13}$ GeV from Eq.~(\ref{Hinf}).
The other slow-roll parameter (\ref{eta}) is approximately 
given by $\eta_V \simeq -(4/3)e^{-\sqrt{6}\phi/(3M_{\rm pl})} 
\simeq -1/N$, so the scalar spectral index (\ref{scalar}) reads
\be
n_s-1 \simeq -\frac{2}{N}\,,
\ee
which is $n_s=0.964$ for $N=55$.
This value is within the 1$\sigma$ observational contour constrained by the Planck CMB data \cite{Planckinf}.

For the massless tensor ($m_g=0$), the power spectrum 
at the end of inflation is given by Eq.~(\ref{Phs}), i.e., 
${\cal P}_T \simeq 2H^2/(\pi^2 M_{\rm pl}^2)$ with 
\be
n_t \simeq -\frac{3}{2N^2}\,,\qquad
r \simeq \frac{12}{N^2}\,.
\ee
For $N=55$ the tensor-to-scalar ratio is
$r \simeq 4.0 \times 10^{-3}$, which is 
much below the current CMB 
bound ($r<0.11$) \cite{Planckinf}.

For the massive tensor, the existence of the parametric
resonance offers the possibility of amplifying tensor 
perturbations to the detectable level in CMB measurements.
In the following, we will numerically compute the 
primordial tensor power spectra both at the onset and 
the end of amplification. 
Since $\phi_{\rm end}$ is of order $M_{\rm pl}$, 
we identify the onset of reheating as $t_i=1/M$.

\subsection{Tensor power spectrum at the onset of reheating}

\begin{figure}
\begin{center}
\includegraphics[height=3.3in,width=3.4in]{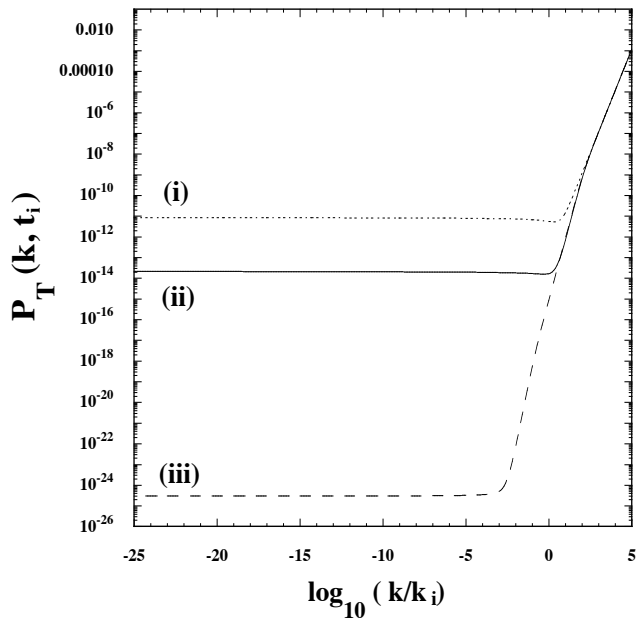}
\end{center}
\caption{\label{fig2}
The primordial tensor power spectra at the onset of reheating ($t=t_i$) 
for the potential (\ref{poten}) with $M=1.3 \times 10^{-5}M_{\rm pl}$.
Each curve corresponds to 
(i) the massless tensor ($m_g^2=0$), 
(ii) the tensor mass squared (2.12) with 
$n=2$, $b=4$, $\lambda=5 \times 10^{-7}$, and 
(iii) the tensor mass squared (2.12) with 
$n=1$, $b=4$, $\lambda=5 \times 10^{-7}$.
The suppression of GWs induced by 
the heavy tensor mass during inflation tends to be 
less significant for larger values of $n$ and $b$.}
\end{figure}

We start to integrate the tensor perturbation Eq.~(\ref{vseq}) 
from the $N=64$ e-folding before the end of inflation.  
The initial condition for the mode functions deep inside the Hubble horizon
is given by Eq~(\ref{WKBsol}).
In Fig.~\ref{fig2}, we show the primordial tensor power spectrum just 
after the end of inflation for three different tensor masses. 

For the standard massless tensor, which corresponds 
to the case (i) in Fig.~\ref{fig2}, the power spectrum of 
GWs which crossed the Hubble horizon before the end of  inflation ($k \lesssim k_i \equiv a_i H_i$) 
is given by ${\cal P}_T \simeq 2H^2/(\pi^2 M_{\rm pl}^2)$, 
while, for $k \gtrsim k_i$, it is given by the 
spectrum (\ref{Phma3}) with the scale-dependence 
${\cal P}_T \propto k^2$.

The plot (ii) in Fig.~\ref{fig2} is the tensor power spectrum 
at $t=t_i$ for the tensor mass squared (\ref{mg1}) with $n=2$, $b=4$, 
and $\lambda=5 \times 10^{-7}$. 
In this case, the tensor mass is in the regime 
$m_g^2 \lesssim H^2$ during most of the inflationary epoch. 
The transition to the regime $m_g^2 \gtrsim H^2$ 
rapidly occurs just before the end of inflation (with the e-folding number 
$N_{f \to i}=0.7$), so the suppression factor $\exp(-3N_{f \to i})$ 
in Eq.~(\ref{Phma1}) is of order 0.1. 
In this case, the wavenumber $k_1$ discussed in Sec.~\ref{trasec} 
corresponds to $k_1 \simeq e^{-N_{f \to i}} k_i \simeq 0.5k_i$.
For the modes $k \ll k_1$, the 
numerically derived power spectrum is nearly 
scale-invariant, whose property is consistent with the analytic result (\ref{Phma1}).
As estimated from Eq.~(\ref{Phma2}), the power 
spectrum has the dependence ${\cal P}_T \propto k^3$ for 
$k_1 \ll k \lesssim k_2 \simeq 10^{5/2}k_i$. 
In the numerical simulation of Fig.~\ref{fig2}, there is an intermediate 
regime around $0.5k_i<k<5k_i$ in which the power spectrum 
has the scale-dependence ${\cal P}_T \propto k^n$ with $0<n<3$. 
The highly blue-tilted spectrum ${\cal P}_T \propto k^3$ arises 
for $k \gtrsim 5k_i$.
For the modes $k \gtrsim k_2 \simeq 10^{5/2}k_i$, the scale 
dependence changes to ${\cal P}_T \propto k^2$ according 
to Eq.~(\ref{Phma3}).

In the case (iii) of Fig.~\ref{fig2}, the power $n$ in the tensor mass is 
smaller than that in the case (ii), so the period of the regime $m_g^2 \gtrsim H^2$
during inflation is longer. Hence the perturbations with the wavenumber 
$k \lesssim 10^{-5/2}k_i$ are subject to stronger suppression. 
In this case, the power spectrum is given by Eq.~(\ref{Phma2}) for the modes 
$10^{-5/2}k_i  \lesssim k  \lesssim 10^{5/2}k_i$ 
and by Eq.~(\ref{Phma3}) for the modes 
$k \gtrsim 10^{5/2}k_i$.

These results show that the suppression of ${\cal P}_T$  tends to be smaller
for $m_g^2(\phi)$ undergoing a faster transition 
from the region $m_g^2 \lesssim H^2$ to the 
region $m_g^2 \gtrsim H^2$ .
This corresponds to the choice of larger values of $n$ and $b$ 
in Eq.~(\ref{mg1}).

\subsection{Tensor power spectrum after inflaton decay}
\label{tendecaysec}

\begin{figure}
\begin{center}
\includegraphics[height=3.2in,width=3.4in]{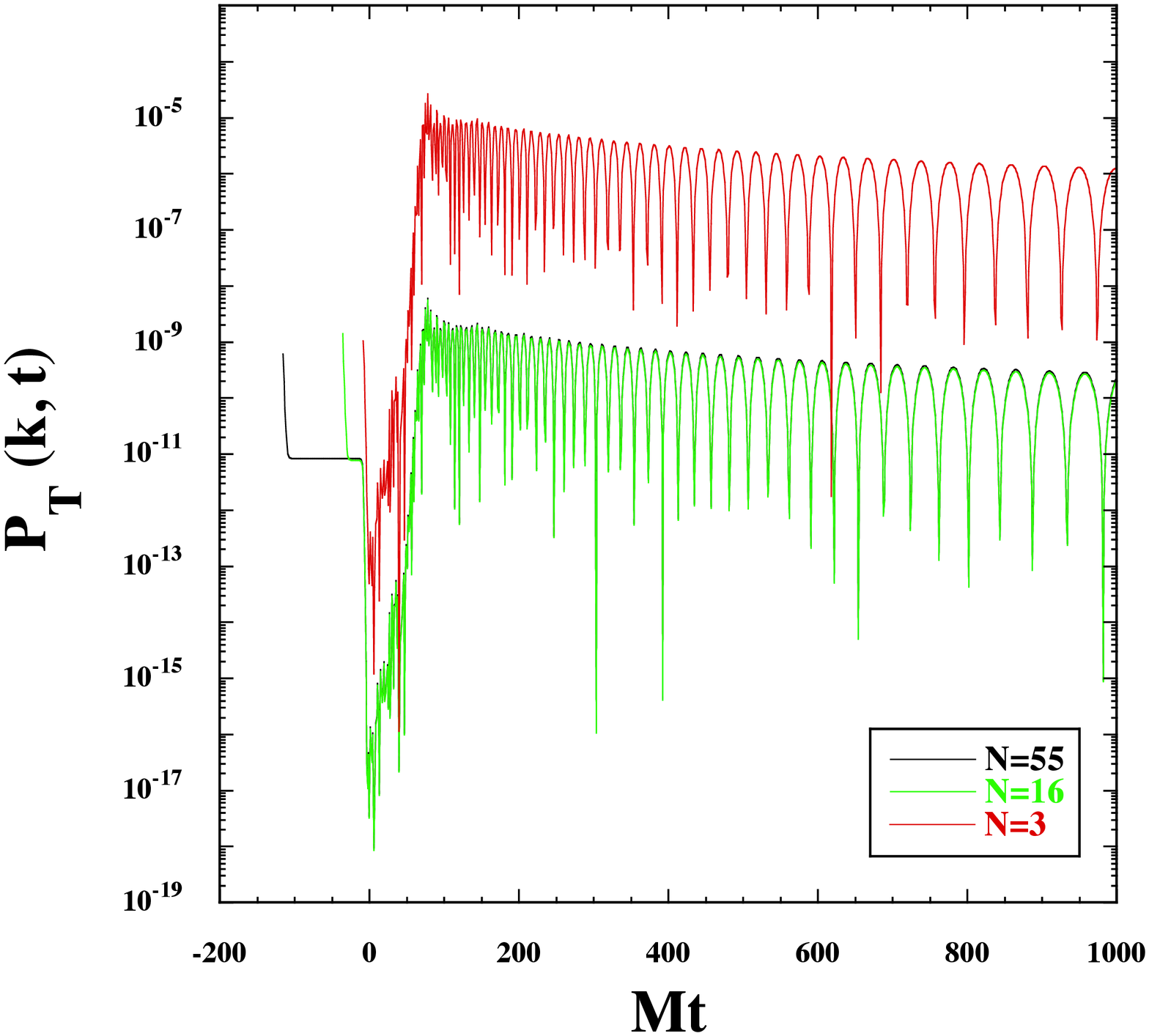}
\end{center}
\caption{\label{fig3}
Evolution of the tensor power spectrum in Starobinsky inflation 
with $M=1.3 \times 10^{-5}M_{\rm pl}$ for the tensor mass 
squared (\ref{mg1}) with $\lambda=4.8 \times 10^{-7}$, 
$n=2$, and $b=2$.
The three curves correspond to the modes which crossed 
the Hubble radius at $N=55,16,3$ e-folding 
before the end of inflation, respectively.}
\end{figure}

During reheating, the tensor perturbation is subject to the broad 
parametric amplification for the coupling constant $\lambda$ 
satisfying the condition (\ref{lamcon}). 
For concreteness of our discussion, in this subsection we consider 
the model parameters $\lambda=4.8 \times 10^{-7}$,
 $M=1.3 \times 10^{-5}M_{\rm pl}$, 
$n=2$, and $b=2$. 
In Fig.~\ref{fig3}, we plot the evolution of the power spectrum ${\cal P}_T$ 
for these model parameters.

The mode which crossed the Hubble horizon
at $N=55$ e-folding before the end of inflation corresponds 
to the perturbation associated with the observation of 
CMB temperature anisotropies. 
As we see in Fig.~\ref{fig3}, this large-scale tensor mode 
is temporarily nearly frozen after horizon crossing ($k<aH$) 
and then it starts to decrease when the tensor mass squared becomes
comparable to $H^2$ during inflation. 
In spite of the decrease of ${\cal P}_T$ by a factor of 
${\cal O}(10^{-6})$ by the end of inflation, the parametric amplification 
of GWs during reheating enhances ${\cal P}_T$ by a factor of ${\cal O}(10^8)$, 
resulting in the final enhancement factor of ${\cal O}(10^2)$. 
For larger $\lambda$, the peak value of ${\cal P}_T$ 
after the amplification ($q \simeq 0.3$) generally increases.

In Fig.~\ref{fig3}, we also show the evolution of ${\cal P}_T$ for the mode 
which crossed the Hubble horizon at the $N=16$ e-folding before the 
end of inflation.  
The evolution of ${\cal P}_T$ during inflation 
and reheating is similar to that for the mode with $N=55$.

The modes that crossed the Hubble radius a few e-folding
($N\lesssim5$) before the end of inflation, namely those in the range
$k_1<k<k_i$, exhibit different evolutionary behavior to the
larger-scale modes. They are not subject to the strong suppression 
during inflation as discussed in Sec.~\ref{trasec}. 
Then, the peak values of ${\cal P}_T$ reached for these 
small-scale perturbations are larger than those for the modes 
which crossed the Hubble horizon earlier ($N \gtrsim 5$), see 
Fig.~\ref{fig3} for $N=3$.
Numerically, we also studied the evolution of GWs for 
smaller-scale modes ($k \gtrsim k_i$) and confirmed that 
the parametric resonance occurs for the 
wavenumber $k<k_{\rm cut}={\cal O}(10^2 k_i)$.

In Fig.~\ref{fig3}, we find that the power spectra 
${\cal P}_T$ start to decrease after reaching their peak values. 
This comes from the fact that the inflaton coherently oscillates 
according to Eq.~(\ref{phiosc}) up to the time 
$t_\Gamma \simeq 1/\Gamma$. 
For $t_i<t<t_\Gamma$, we take 
the time average of the tensor mass squared over oscillations, 
such that $\langle m_g^2(\phi) \rangle 
\simeq \langle \lambda \phi^2 \rangle \simeq 4\lambda M_{\rm pl}^2/(3M^2t^2)$. 
Provided that  $\langle m_g^2(\phi) \rangle$ dominates over the gradient term 
$k^2/a^2$, Eq.~(\ref{vseqt}) reduces to 
\be
\ddot{v}+\frac{2}{t} \dot{v}
+\frac{4\lambda M_{\rm pl}^2}
{3M^2t^2}v \simeq 0\,,
\ee
where we used the fact that the scale factor evolves as 
$a \propto t^{2/3}$ for $t_i<t<t_{\Gamma}$.
The solution to this equation is given by 
\be
v\propto t^{-1/2\pm i\Omega}\,,
\ee
where $\Omega=(48\lambda M_{\rm pl}^2/M^2-9)^{1/2}/6>0$
under the condition (\ref{lamcon}), i.e., $\lambda M_{\rm pl}^2/M^2\gg1$.
So the amplitude of GWs decreases as 
$|v| \propto t^{-1/2}$.  
Thus, for large-scale modes satisfying the condition 
$k^2/a^2< \langle m_g^2(\phi) \rangle$ until the time $t_{\Gamma}$, 
the amplitude of ${\cal P}_T$ decreases 
as $\langle {\cal P}_T \rangle \propto t^{-1}$ after reaching its peak value. 
This behavior is confirmed in the numerical results shown in Fig.~\ref{fig3}.

For small-scale modes, it happens that the gradient term 
$k^2/a^2$ ($\propto t^{-4/3}$) gets larger than 
$\langle m_g^2(\phi) \rangle$ ($\propto t^{-2}$)
during the time interval $t_i<t<t_\Gamma$. 
The time $t_k$ at which $k^2/a^2$ becomes equivalent to 
$\langle m_g^2(\phi) \rangle$ can be estimated as 
\be
\left( \frac{t_k}{t_i} \right)^{2/3} 
\simeq 3\lambda \left( \frac{M_{\rm pl}}{M}
\right)^2 \left( \frac{k_i}{k} \right)^2\,.
\ee
For example, for $M=1.3 \times 10^{-5}M_{\rm pl}$ 
and $\lambda=4.8 \times 10^{-7}$, we find
$t_k \simeq 800t_i$ for $k=10k_i$.
During the time interval $t_i<t<t_k$ the amplitude of GWs 
decreases as $\langle {\cal P}_T \rangle \propto t^{-1}$, 
while, for $t>t_k$, $\langle m_g^2(\phi) \rangle$ gradually 
becomes negligible relative to the term 
$k^2/a^2$. Provided that $\langle m_g^2(\phi) \rangle \ll k^2/a^2$ 
the power spectrum evolves as 
$\langle {\cal P}_T \rangle \propto a^{-2} \propto t^{-4/3}$, 
thus decreasing faster than that during $t_i<t<t_k$. 

\begin{figure}
\begin{center}
\includegraphics[height=3.2in,width=3.4in]{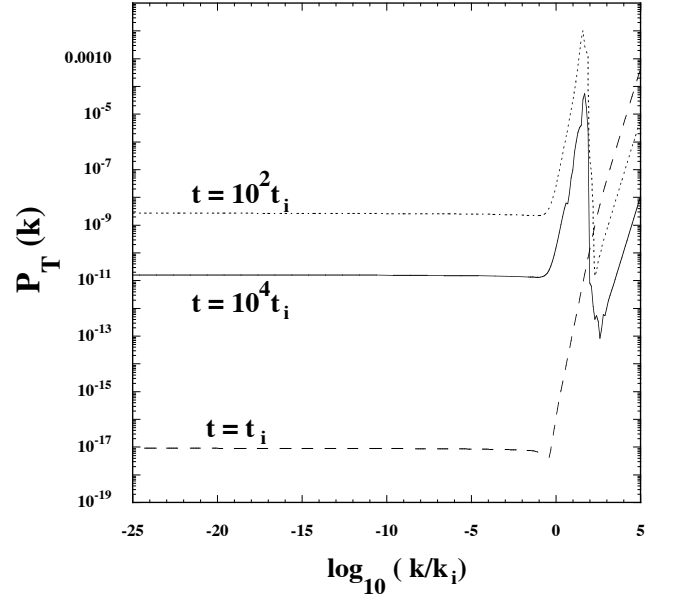}
\end{center}
\caption{\label{fig4}
The tensor power spectra at $t=10^2t_i$ and $t=10^4t_i$
for the potential (\ref{poten}) with 
$M=1.3 \times 10^{-5}M_{\rm pl}$ in the presence of 
the tensor mass squared (\ref{mg1})
with $\lambda=4.8 \times 10^{-7}$, $n=2$, and $b=2$. 
We also show the spectrum at the onset of 
reheating ($t=t_i$). After the amplitude of perturbations 
reaches the maximum around the time $t \simeq 80t_i$, 
it decreases until the moment $t_{\Gamma}$ 
at which the inflaton decays to the radiation.}
\end{figure}

In Fig.~\ref{fig4}, we plot the tensor power spectra at three 
different epochs ($t=t_i, 10^2t_i, 10^4t_i$) 
for the same model parameters used in Fig.~\ref{fig3}.
The modes which crossed the Hubble horizon before the 
onset of reheating are subject to the suppression during inflation, 
but the parametric resonance leads to the growth 
of ${\cal P}_T$ with a factor of ${\cal O}(10^8)$ 
by the time $t_{\rm max} \simeq 80t_i$ (at which ${\cal P}_T$ 
reaches the maximum value).
The spectrum of the modes $k \lesssim k_1(\approx 0.3k_i)$ is nearly 
scale-invariant after the parametric amplification.
The GWs with the wavenumber 
$k_i \lesssim k \lesssim 10^2 k_i$
have the highly blue-tiled spectrum (\ref{Phma2}) at the onset of reheating ($t=t_i$).
These modes are excited by the parametric resonance, 
but the perturbations with the wavenumber $k>k_{\rm cut} \approx 10^2k_i$ are not subject to the amplification.
The power spectrum ${\cal P}_T(k)$ at $t=10^2t_i$ has a peak 
around $k \simeq 40k_i$. For the modes $k \gtrsim 10^2k_i$, the gradient 
term $k^2/a^2$ dominates over $m_g^2(\phi)$, 
so the power spectrum at $t=10^2t_i$ has the dependence 
${\cal P}_T \propto k^2$.

In summary, during the time interval $t_{\rm max}<t<t_{\Gamma}$, 
the amplitude decreases as 
$\langle {\cal P}_T \rangle \propto t^{-1}$ for the modes 
$k^2/a^2 \ll \langle m_g^2(\phi) \rangle$, while 
$\langle {\cal P}_T \rangle \propto t^{-4/3}$ for the modes
$k^2/a^2 \gg \langle m_g^2(\phi) \rangle$. 
This behavior can be confirmed by comparing the two spectra 
at $t=10^2t_i$ and $t=10^4 t_i$ shown in Fig.~\ref{fig4}. 
Depending on the time $t_{\Gamma}$ after which 
$\phi$ decreases exponentially and hence
$m_g^2$ becomes completely negligible in Eq.~(\ref{vseqt}), 
the amplitude of today's GW power spectrum  
is different (as we will discuss in Sec.~\ref{GWbackground}).
We regard ${\cal P}_T(t_{\Gamma})$ as the primordial tensor 
power spectrum at which the super-horizon
tensor perturbations ($k/a<H$) are frozen.
For $t>t_{\Gamma}$ the GW  evolves as a 
massless field, i.e., 
${\cal P}_T (t)={\rm constant}$ for $k/a<H$ and 
$\langle {\cal P}_T (t) \rangle \propto a^{-2}$ for $k/a>H$ \cite{Kuro08}.

The power spectrum ${\cal P}_T(t_{\Gamma})$ 
tends to be smaller if we choose smaller values of $\Gamma$. 
If the inflaton decay occurs at the time $10^2t_i$ in the 
numerical simulation of Fig.~\ref{fig4}, then ${\cal P}_T(t_{\Gamma})$ is 
of order $10^{-9}$ for the modes $k \lesssim k_i$. 
This case is excluded from the Planck CMB bound 
${\cal P}_T(t_{\Gamma})<2.4 \times 10^{-10}$ \cite{Planckinf}.
For $t_{\Gamma}=10^4t_i$ the amplitude of 
${\cal P}_T(t_{\Gamma})$ for large-scale modes
is of order $10^{-11}$, so this case is consistent 
with the CMB measurements.
For the coupling $\lambda$ used in 
Fig.~\ref{fig4}, the decay constant is constrained to be 
$\Gamma \lesssim 10^{-3}M \simeq 10^{-8}M_{\rm pl}$. 
It is possible to realize a larger maximum value of 
${\cal P}_T$ if we choose a larger $\lambda$, 
in which case the upper bound on 
$\Gamma$ constrained from CMB tends to be smaller. 

The peaks of ${\cal P}_T (k)$ appearing in Fig.~\ref{fig4}
affect today's energy density spectrum 
$\Omega_{\rm GW}$ of the 
GW background. In Sec.~\ref{GWbackground}, we will 
explicitly compute $\Omega_{\rm GW}$ for the model 
discussed above.

\section{Low-scale inflation with the 
$\dot{\phi}$-dependent tensor mass}
\label{lowsec}

In this section, we study the low-scale inflationary scenario in which the slow-roll 
parameter $\epsilon$ is much smaller than $10^{-4}$ during most stage of inflation. 
As an explicit example, we consider a very flat potential 
$V(\phi) \simeq V_0$ for $\phi \gtrsim M_{\rm pl}$ and 
then the rapid transition to the potential 
\be
V(\phi)=\frac12 M^2 \phi^2\,,
\label{massivepo}
\ee
occurs around $\phi \approx M_{\rm pl}$. 
This type of transition often arises in the context of string 
inflation \cite{sinflation}.
We consider the $\dot{\phi}$-dependent tensor mass 
squared (\ref{mg2}) to discuss the dynamics of GWs 
during inflation and reheating.

We consider the case in which the field value at the end of inflation for the potential (\ref{massivepo}) is determined by 
the condition $\epsilon_V=1$, i.e., 
$\phi_{\rm end}=\sqrt{2}M_{\rm pl}$ with 
$\dot{\phi}_{\rm end}^2=(2/3)M^2M_{\rm pl}^2$.
The Hubble parameter $H_i$ at the onset of reheating ($t=t_i$) obeys the relation 
$3M_{\rm pl}^2H_i^2=
\dot{\phi}_{\rm end}^2/2+M^2 \phi_{\rm end}^2/2$, and hence
\be
H_i \simeq \frac{2}{3} M\,.
\label{Hi}
\ee
The Hubble parameter $H_{\rm inf}$ related to the 
observed CMB temperature anisotropies can be regarded as the same 
order as $H_i$, so Eq.~(\ref{Hinf}) gives the following estimate
\be
M \approx \sqrt{\epsilon} \times 10^{15}~{\rm GeV}\,.
\ee
For inflation satisfying the condition $\epsilon \ll 10^{-4}$, the mass 
scale $M$ is much smaller than $10^{13}$~GeV with the 
tensor-to-scalar ratio $r=16\epsilon \ll 10^{-3}$. 
For the massless tensor, there is almost no hope for 
the detection of primordial tensor modes in both CMB and 
direct detection measurements of GWs, but 
this is not the case for the massive tensor.

In this section, for concreteness, we adopt
the model parameters $\mu=1.523 \times 10^4$ and $M=10^8$ GeV,
which gives $\phi_{\rm end}={\cal O}(M_{\rm pl})$.
We note, however, that it is also possible to realize low-scale inflation 
with $\phi_{\rm end} \ll M_{\rm pl}$, in which case 
$H_i \ll M$. In fact, the potential $V(\phi)=V_0(1
-e^{-\alpha \phi/M_{\rm pl}})^2$ of the $\alpha$-attractor 
model leads to $\phi_{\rm end} \ll M_{\rm pl}$ for 
$\alpha \gg 1$. As we discussed in Sec.~\ref{parasec}, however, 
the broad parametric resonance tends to be less efficient for such smaller 
values of $\phi_{\rm end}$. 
We need to choose a larger coupling $\mu$ to realize 
the parametric resonance comparable to the case 
$\phi_{\rm end}={\cal O}(M_{\rm pl})$, but in such cases 
the ratio $m_g^2/H^2 \simeq 2\mu \epsilon$ during inflation 
gets larger. This leads to the earlier entry to the massive 
regime $m_g^2>H^2$, so the GWs are subject to stronger 
suppression during inflation. 
Hence we will focus on the low-scale inflation where the 
transition to the reheating stage given by 
the potential (\ref{massivepo}) occurs around 
$\phi \approx M_{\rm pl}$. In this case, we can identify 
the time $t_i$ at the onset of reheating as $z_i=Mt_i=1$.

{}From Eq.~(\ref{qmu}) the resonance parameter $q$ 
is much larger than 1 at $t=t_i$ for the coupling 
$\mu \gg 1$, in which case the broad parametric 
resonance occurs by the coherent 
oscillation of the inflaton around $\phi=0$. 
In low-scale inflation the slow-roll parameter $\epsilon$ 
during inflation is very much smaller than 1, 
so even the large coupling 
with $\mu \gg 1$ allows one to satisfy the condition 
$m_g^2/H^2=2\mu \epsilon \ll 1$.
Since $\epsilon$ rapidly grows from a tiny value to unity
during a very short period around the end of inflation, 
it is possible to avoid the suppression of GWs induced by the 
growth of $m_g^2$. 

We recall that, for the modes $k<k_1$, 
the tensor power spectrum 
at the end of inflation is estimated as Eq.~(\ref{Phma1}).
Since $m_g^2/H^2$ and $\epsilon$ are much smaller 
than 1 during most stage of low-scale inflation, 
the power spectrum at $t=t_i$ reduces to  
\begin{eqnarray}
{\cal P}_T (t_i) &\simeq& 
\frac{2H_i^2}
{\pi^2 M_{\rm pl}^2}
\frac{m_g(t_f)}{m_g(t_i)}
e^{-3N_{f \to i}} \nonumber \\ 
&\simeq& 3.4 \times 10^{-8} \epsilon_{\rm CMB}\, 
\frac{m_g(t_f)}{m_g(t_i)}
e^{-3N_{f \to i}} \,,
\label{Phti}
\end{eqnarray}
where we have employed the approximations that 
$H(t_f)$ is equivalent to $H_i$ and that the exponential 
factor in the second line of Eq.~(\ref{Phma1}) equals to 1. 
Note that $\epsilon_{\rm CMB}$ is the slow-roll parameter 
associated with the perturbation relevant to the CMB 
observations. There is the suppression factor $e^{-3N_{f \to i}}$ 
in Eq.~(\ref{Phti}). 
As long as inflation ends shortly after $m_g^2$ 
crosses $H^2$, the suppression induced by the massive 
tensor is not so significant. 
For $\epsilon \ll 10^{-4}$ the initial power spectrum 
(\ref{Phti}) is much smaller than the order of $10^{-12}$,
but the oscillating tensor mass makes it possible to amplify 
the GWs to the detectable level of CMB observations.

The power spectrum at $t=t_i$ for the modes 
$k_1<k<k_2$ is given by Eq.~(\ref{Phma2}), where 
$k_1$ may be expressed as
\be
\frac{k_1}{a_iH_i}=\frac{a(t_f)H(t_f)}{a_iH_i}
\simeq e^{-N_{f \to i}}\,,
\label{k1est}
\ee
where we used the approximation $H(t_f) \simeq H_i$.
For the modes $k>k_1$, 
we numerically integrate Eq.~(\ref{vseqt}) by using the 
initial condition (\ref{vkin}) at $t=t_i$. 
As for the modes $k<k_1$, since they are amplified with the 
$k$-independent growth rate from the initial scale-invariant 
spectrum (\ref{Phti}), the shape of the spectrum remains the
same as the original one, that is, it is almost scale-invariant.

\begin{figure}
\begin{center}
\includegraphics[height=3.2in,width=3.4in]{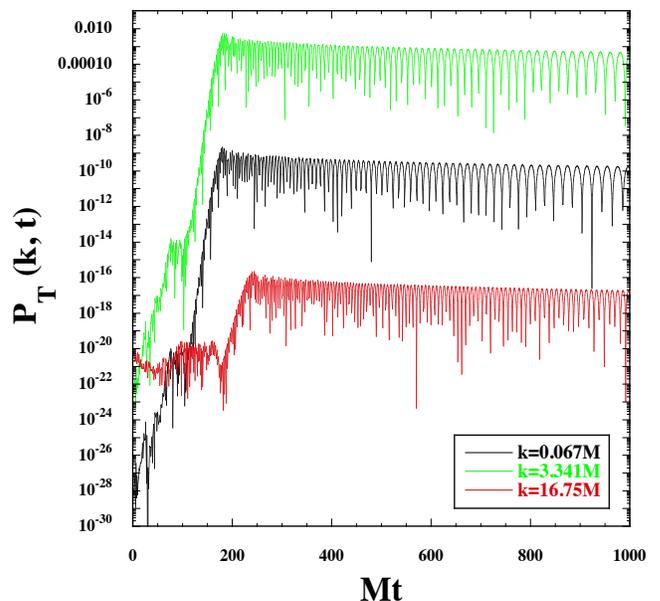}
\end{center}
\caption{\label{fig5}
Evolution of the tensor power spectrum during 
reheating in low-scale inflation with 
$M=4.1 \times 10^{-11}M_{\rm pl}=10^8$~GeV for 
the tensor mass squared (\ref{mg2}) with 
$\mu=1.523 \times 10^{4}$.
We integrate Eq.~(\ref{vseqt}) from the onset of reheating 
($a_i=1$) with the initial conditions $\phi_i=\sqrt{2}M_{\rm pl}$ 
and $\dot{\phi}_i=-\sqrt{2/3}MM_{\rm pl}$.
Each curve corresponds to the evolution of ${\cal P}_T$  
for $k=0.067M$, $3.341M$, and $16.75M$, respectively.}
\end{figure}

\begin{figure}
\begin{center}
\includegraphics[height=3.2in,width=3.4in]{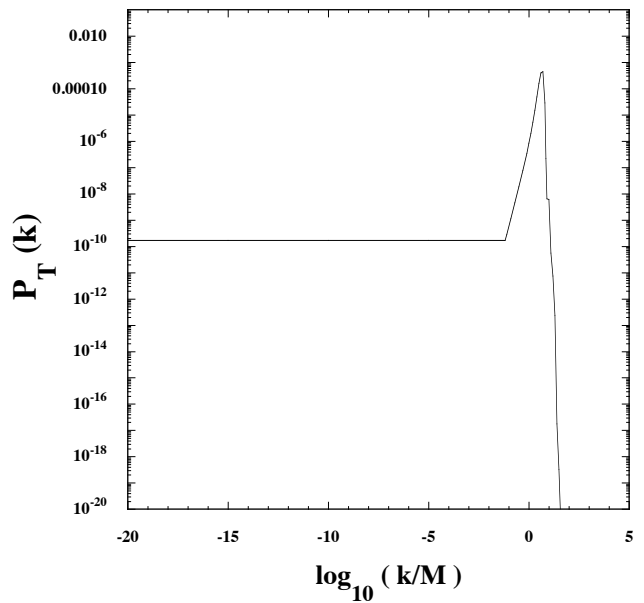}
\end{center}
\caption{\label{fig6}
The tensor power spectrum 
at $t=10^3/M$ in low-scale 
inflation with the tensor mass squared (\ref{mg2})  
for the same model parameters used in Fig.~\ref{fig5}.}
\end{figure}

In Fig.~\ref{fig5}, we plot the evolution of ${\cal P}_T$ 
from the end of inflation for three different values of $k$,
for the assumed values of the parameters, $\mu=1.523 \times 10^4$ and $M=10^8$ GeV.
In this case, the suppression factor 
$e^{-N_{f \to i}}$ is equal to $0.1$, 
so that the wavenumber (\ref{k1est}) is given by 
$k_1/(a_iH_i)=0.1$, i.e., 
$k_1=0.067M$ by setting $a_i=1$. 
In this case, the slow-roll parameter $\epsilon_{\rm CMB}$ 
is of order $10^{-14}$ from Eq.~(\ref{Hinf}) and the tensor mass 
grows to the value of order $m_g/H_i=\sqrt{2\mu}
={\cal O}(10^2)$ by the end of inflation. 
{}From Eq.~(\ref{Phti}) we see that the power spectrum for the modes 
$k<k_1$ is as small as ${\cal P}_T (t_i) \approx 10^{-27}$
at the onset of reheating.

As we see in Fig.~\ref{fig5}, the power spectrum ${\cal P}_T$ 
for the mode $k=k_1=0.067M$ is subject to a strong amplification 
from the initial value of order $10^{-27}$ to the 
maximum value of around $10^{-9}$.  
Analogous to Eq.~(\ref{zf}) the end of amplification can be estimated as
$Mt_e \simeq 1.5 \sqrt{\mu} \simeq 180$, which is in good 
agreement with the numerical results shown in Fig.~\ref{fig5}.
The power spectrum starts to decrease after reaching 
its maximum value around $t=t_f$.
This is attributed to the fact that the tensor mass squared 
averaged over oscillations evolves as 
$\langle m_g^2(\dot{\phi}^2) \rangle \propto t^{-2}$ 
up to the time $t_{\Gamma}=1/\Gamma$. 
For the wavenumbers satisfying the condition 
$k^2/a^2<\langle m_g^2(\dot{\phi}^2) \rangle$ until $t=t_{\Gamma}$, 
${\cal P}_T$ decreases in proportion 
to $t^{-1}$. For $t>t_{\Gamma}$ the inflaton decays 
to the radiation, so the effect of the tensor mass 
on the evolution of ${\cal P}_T$ becomes negligible.

For the wavenumbers in the range $k_1<k<k_2$, 
the power spectrum at $t=t_i$ is given by Eq.~(\ref{Phma2}).
Provided that the condition 
$k^2/a_i^2 \ll \mu \dot{\phi}_i^2/M_{\rm pl}^2$ is satisfied, 
these modes are amplified by approximately the same factor
as those in the range $k<k_1$. Hence the shape of the initial 
blue-tilted spectrum ${\cal P}_T (t_i) \propto k^3$ remains
the same, and the resulting amplitude of GWs starts to increase 
for $k$ larger than $k_1$. 
Indeed, the numerical result given in Fig.~\ref{fig5} shows 
that the maximum value of ${\cal P}_T$ reached for 
$k=3.34M$ is larger than that for  $k=k_1=0.067M$. 
For the wavenumber $k \gtrsim 10M$, the parametric excitation
of GWs tends to be less efficient, see the evolution of ${\cal P}_T$ 
for the mode $k=16.75M$ in Fig.~\ref{fig5}.
The parametric resonance does not occur for the modes 
$k^2/a_i^2 > \mu \dot{\phi}_i^2/M_{\rm pl}^2$, which 
translates to the condition 
$k \gtrsim k_{\rm cut} \approx \sqrt{\mu}M \approx 10^2M$.

In Fig.~\ref{fig6}, we plot the power spectrum ${\cal P}_T(k)$ at 
$t=10^3/M$ for the same values of $\mu$ and $M$ as those 
used in Fig.~\ref{fig5}. As explained above, ${\cal P}_T(k)$ 
starts to increase for the modes $k>k_1=0.067M$ and it reaches 
the peak value around $k=3.3M$.
There is a sharp drop-down of ${\cal P}_T(k)$ 
for the modes $k \gtrsim 10M$.
As we discussed in Sec.~\ref{tendecaysec}, the final amplitude of 
${\cal P}_T (k)$ depends on the time when the inflaton decays to radiation. 
If the decay occurs earlier (later) than $t=10^3/M$, 
the resulting amplitude of GWs tends to be larger (smaller) 
than ${\cal P}_T(k)$ shown in Fig.~\ref{fig6}.

\section{Spectrum of the gravitational wave background}
\label{GWbackground}

In this section, we compute the spectrum of the GW background generated in two models discussed in Secs.~\ref{Stasec} and \ref{lowsec}. The intensity of the GW background is conventionally defined as 
\be
\Omega_{\rm GW} \equiv 
\frac{1}{\rho_c} \frac{d \rho_{\rm GW}}
{d \ln k}\,,
\label{OmegaGW0}
\ee
where $\rho_c=3M_{\rm pl}^2 H^2$ is the critical density of 
the Universe and $\rho_{\rm GW}$ is the energy density 
of GWs \cite{Allen,Maggiore}. 
The GW intensity can be expressed as \cite{Yokoyama,Kuro11,Kuro14}
\be
\Omega_{\rm GW}(k,t)=\frac{1}{12} 
\left( \frac{k}{aH} \right)^2 {\cal P}_T (k, t)\,,
\label{OmegaGW}
\ee
where ${\cal P}_T (k, t)$ is defined by Eq.~(\ref{powerspe}).

After the inflaton decays to radiation at 
$t=t_{\Gamma}$, the GWs evolve as the standard massless tensor perturbation. 
For the modes outside the Hubble radius ($k<aH$) at time $t_{\Gamma}$, 
the tensor perturbation is frozen until the second horizon crossing 
(labelled by ``sh''). After the second horizon crossing, the GWs 
evolve as $v \propto a^{-1}e^{\pm ik \tau}$.
Hence the present power spectrum ${\cal P}_T (k, t_0)$ 
is related to the one at time $t_{\Gamma}$, as 
\be
{\cal P}_T (k, t_0)={\cal P}_T (k, t_{\Gamma}) 
\left( \frac{a_{\rm sh}}{a_0} \right)^2\,,
\ee
where the subscript ``$0$" denotes today's value. 
In the Universe where the scale factor evolves as 
$a \propto t^p$, with $p$ being a constant, 
the scale factor at the second horizon crossing 
has the $k$-dependence $a_{\rm sh} \propto k^{p/(p-1)}$.
For the primordial power spectrum ${\cal P}_T (k, t_{\Gamma})$ 
with the spectral index $\tilde{n}_t$, i.e., 
${\cal P}_T (k, t_{\Gamma}) \propto k^{\tilde{n}_t}$, 
today's intensity (\ref{OmegaGW}) of the 
GW background has the $k$-dependence,
\be
\Omega_{\rm GW} (k, t_0) 
\propto k^{\tilde{n}_t+2+2p/(p-1)}\,.
\ee
Hence $\Omega_{\rm GW} (k, t_0) \propto k^{\tilde{n}_t}$ 
in the radiation era ($p=1/2$) and 
$\Omega_{\rm GW} (k, t_0) \propto k^{\tilde{n}_t-2}$ 
in the matter era ($p=2/3$). 
For $\tilde{n}_t=0$,  
we have the scale-invariant spectrum 
$\Omega_{\rm GW} (k, t_0) \propto k^{0}$ for 
$p=1/2$ and the red-tilted spectrum 
$\Omega_{\rm GW} (k, t_0) \propto k^{-2}$ for 
$p=2/3$, respectively. 
The existence of peaks in ${\cal P}_T (k, t_{\Gamma})$ 
seen in Figs.~\ref{fig4} and \ref{fig6} should give rise to 
specific features in $\Omega_{\rm GW} (k, t_0)$ 
(as we will discuss below). 

The position of peaks in ${\cal P}_T (k, t_{\Gamma})$ is related to  
the wavenumber $k_i=a_iH_i$ which crossed the Hubble radius 
at the end of inflation. Assuming that the entropy before the inflaton decay 
is conserved in the photon and neutrino background today 
(temperature $T_0$), the scale factor $a_{\Gamma}$ at time 
$t_{\Gamma}$ is given by \cite{Dai,Kuro14}
\be
a_{\Gamma}=a_0 \left( \frac{11}{43}g_{*} \right)^{-1/3} 
\frac{T_0}{T_{\Gamma}}\,,
\ee
where $T_{\Gamma}$ is the reheating temperature 
given by Eq.~(\ref{Tg}). 
During the time interval $t_i<t<t_{\Gamma}$, the 
energy density of the Universe decreases as 
$\rho \propto a^{-3}$, so we have
$a_i=a_{\Gamma}(\rho_{\Gamma}/\rho_i)^{1/3}$, where 
$\rho_i \simeq 3M_{\rm pl}^2H_i^2$ and 
$\rho_{\Gamma} =\pi^2 g_*T_{\Gamma}^4/30$.
On using the fact that today's temperature is
$T_0 \simeq 9.64 \times 10^{-32}M_{\rm pl}$, 
the frequency $f_i=k_{i}/(2\pi)=a_iH_i/(2\pi)$ can 
be estimated as 
\begin{align}
f_i &\simeq 3.0 \times 10^{10} 
\left(\frac{g_*}{106.75}\right)^{-1/12}
\left( \frac{H_i}{M_{\rm pl}} 
\right)^{1/3} \left( \frac{\Gamma}{M_{\rm pl}} 
\right)^{1/6}{\rm Hz}
\cr
&=3.0 \times 10^{10} 
\left(\frac{g_*}{106.75}\right)^{-1/12}
\left( \frac{H_i}{M_{\rm pl}}\right)^{1/2}
\left( \frac{\Gamma}{H_i}\right)^{1/6}{\rm Hz}.
\label{fi}
\end{align}
One observes that $f_i$ can become very small and 
falls in the A-LIGO band for a very small $H_i$, 
say $H_i/M_{\rm pl}\sim 10^{-18}$.
In the following, we assume that the relativistic degree of freedom 
is equivalent to $g_*=106.75$ at $t=t_{\Gamma}$. 

\subsection{Starobinsky inflation}

In the Starobinsky model with the massless tensor, 
the tensor-to-scalar ratio $r$ is of the order $10^{-3}$ at
the CMB scale, which is much below the current observational bound ($r<0.11$). 
On the other hand, the parametric resonance induced by the 
massive tensor may make it possible to detect the GWs
in CMB measurements.

\begin{figure}
\begin{center}
\includegraphics[height=3.5in,width=3.5in]{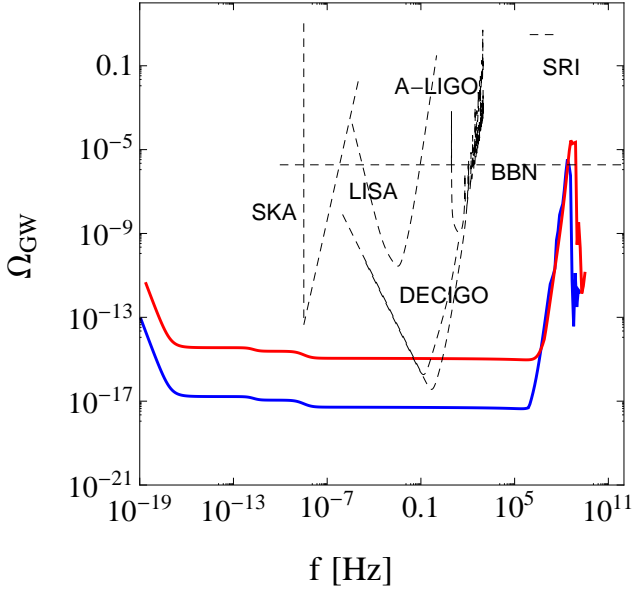}
\end{center}
\caption{\label{fig7}
Today's spectral energy 
density  $\Omega_{\rm GW}$ of the GW background 
versus the frequency $f=k/(2\pi)$ in the Starobinsky model with the $\phi$-dependent tensor mass 
squared (\ref{mg1}) for $M=1.3 \times 10^{-5}M_{\rm pl}$, 
$\lambda=4.785 \times 10^{-7}$, $n=2$, and $b=2$.
We also show the sensitivity curves 
for DECIGO \cite{DECIGO}, upgraded DECIGO \cite{Yagi}, 
A-LIGO \cite{ALIGO}, LISA \cite{LISA}, 
SKA \cite{SKA}, and SRI \cite{SRI}, together 
with the upper bound from the BBN \cite{Cabass:2015jwe}.
For a recent BBN status report, see, e.g., 
Ref.~\cite{Cyburt:2015mya}.
Each line corresponds to the case in which the inflaton 
decay to the radiation occurs at $t=10^2/M$ (red) and 
$t=10^4/M$ (blue).}
\end{figure}

In Fig.~\ref{fig7}, we plot today's spectral energy 
density $\Omega_{\rm GW}$ versus 
the frequency $f$ in Starobinsky inflation with 
the $\phi$-dependent tensor mass squared (\ref{mg1}) 
for $M=1.3 \times 10^{-5}M_{\rm pl}$, 
$\lambda=4.785 \times 10^{-7}$, $n=2$, and $b=2$.
In this case, the tensor power spectrum ${\cal P}_T$ 
relevant to the CMB grows to the amplitude 
$7 \times 10^{-10}$ around the time $t=10^2/M$.
The top (red) line in Fig.~\ref{fig7} corresponds to the case in which 
the inflaton decays to the radiation at time $t=10^2/M$. 
The decay constant in this case is given by $\Gamma=10^{-2}M$ 
with $H_i \simeq 0.3M$, so we have 
$f_i \simeq 3 \times 10^7$ Hz from Eq.~(\ref{fi}). 
Since the primordial spectrum ${\cal P}_T(k)$ at $t=10^2/M$ 
has a peak around $k=40k_i$, this structure is inherited to 
$\Omega_{\rm GW}(f, t_0)$ with the peak around the frequency 
$f=40f_i \simeq 10^9$ Hz.

Although the peak position of $\Omega_{\rm GW}(f, t_0)$ is
at a frequency much higher than the ranges relevant to the detection sensitivities of A-LIGO and DECIGO, we find 
there is a range of frequencies around $f=0.1$ Hz in which 
the theoretical line is within the DECIGO detection range.
The GWs around $f=0.1$ Hz are the ones from the nearly 
scale-invariant primordial perturbations whose horizon re-entry
occurs in the radiation era, so $\Omega_{\rm GW}(f,t_0)$ is almost scale-invariant 
around those frequencies.

For the decay constant $\Gamma=10^{-2}M$, the maximum value of $\Omega_{\rm GW}(f, t_0)$ exceeds the upper bound constrained from the Big Bang Nucleosynthesis (BBN) \cite{Cabass:2015jwe}. 
In the following, we will explain how the BBN places 
bounds on the GW amplitude in considerable detail.
Neutrinos contribute to the energy density of radiation, and 
their energy density is written using the effective number of neutrinos $N_{\rm eff}^\nu$ 
as $\rho_\nu=(7/8)(4/11)^{4/3}N_{\rm eff}^\nu\,\rho_\gamma$, where $\rho_\gamma$ 
is the energy density of photons.  
The GW contribution to the radiation energy density 
at the time of BBN can be effectively described by 
$N_{\rm eff} = N_{\rm eff}^\nu + N_{\rm eff}^{\rm GW}$ with
\beq
N_{\rm eff}^{\rm GW}=\frac{8}{7}\left(\frac{g_{*,s}(T=1~{\rm MeV})}{g_{*,s}(T_0)}\right)^{4/3}\frac{\rho_{{\rm GW},0}}{\rho_{\gamma,0}}
\nonumber \\
=\frac{h^2}{5.6\times 10^{-6}}\int d(\ln f)~ \Omega_{\rm GW}(f), 
\eeq
where $g_{*,s}$ is the effective number of degrees of freedom for entropy and we adopted the standard values $g_{*,s}(T_0)=3.91$ and $g_{*,s}(T=1~{\rm MeV})=10.75$.
In the second step, we have used the definition of 
$\Omega_{\rm GW}$ in Eq.~(\ref{OmegaGW0}) and 
today's density parameter of photons 
$\Omega_{\gamma,0}=\rho_{\gamma,0}
/(3M_{\rm pl}^2 H_0^2) \simeq 2.47 \times 10^{-5}h^{-2}$, 
where $H_0=100\,h$ km~s$^{-1}$~{\rm Mpc}$^{-1}$ 
is today's Hubble expansion rate. 
Thus, using the upper limit $N_{\rm eff}<3.2$ and the standard prediction of the effective number of neutrinos  
$N_{\rm eff}^\nu=3.045$ \cite{Salas}, we can obtain the upper limit on the GW amplitude as \cite{Maggiore,Cabass:2015jwe}
\beq
\int^{10^{10}{\rm Hz}}_{10^{-10}{\rm Hz}}d(\ln f)~ \Omega_{\rm GW}(f) 
&<& \frac{5.6\times 10^{-6}}{h^2}(N_{\rm eff}-N_{\rm eff}^\nu) \nonumber \\
&<& 1.9\times 10^{-6}\,,
\label{Hi2}
\eeq
where the reduced Hubble constant 
$h\simeq 0.6763$ \cite{Ade:2015xua} 
has been used in the second line.

For smaller $\Gamma$, the overall amplitude of 
$\Omega_{\rm GW}(f, t_0)$ decreases with the 
shift of the peak position toward smaller frequencies, 
see the blue line in Fig.~\ref{fig7} for the decay constant 
$\Gamma=10^{-4}M$. 
Provided that $\Gamma<10^{-4}M$,  
the model with the coupling 
$\lambda=4.785 \times 10^{-7}$ is within the BBN bound. 
For such a decay constant, the theoretical line is below the 
sensitivity region of DECIGO.
For smaller $\lambda$ the parametric resonance is less efficient, so the resulting amplitude of 
$\Omega_{\rm GW}(f, t_0)$ gets smaller.
In such cases, the BBN constraint can be satisfied for 
larger $\Gamma$.
However, to satisfy the BBN bound, the model with $n=2$ and
$b=2$ seems to predict the value of 
$\Omega_{\rm GW}(f, t_0)$ too small to be in the DECIGO sensitivity region. 
There may be some cases with different values of 
$n$ and $b$ in which the theoretical line is within the 
DECIGO detection range while satisfying the BBN 
bound, but the parametric resonance is typically too
efficient to give rise to a sharp peak which overshoots the
BBN upper limit.

\subsection{Low-scale inflation}

In low-scale inflation discussed in Sec.~\ref{lowsec}, 
the tensor power spectrum relevant to CMB measurements is very small 
at the end of inflation, but the parametric resonance driven by the
 tensor mass squared (\ref{mg2}) can amplify the GWs to ${\cal P}_T$ 
 larger than $10^{-10}$.
Moreover, since the Hubble parameter $H_i$ at the onset of reheating 
and the decay constant 
$\Gamma$ are much smaller than those in Starobinsky inflation, 
the frequency $f_i$ relevant to the peak position of 
$\Omega_{\rm GW}$ shifts toward smaller values.

\begin{figure}
\begin{center}
\includegraphics[height=3.5in,width=3.5in]{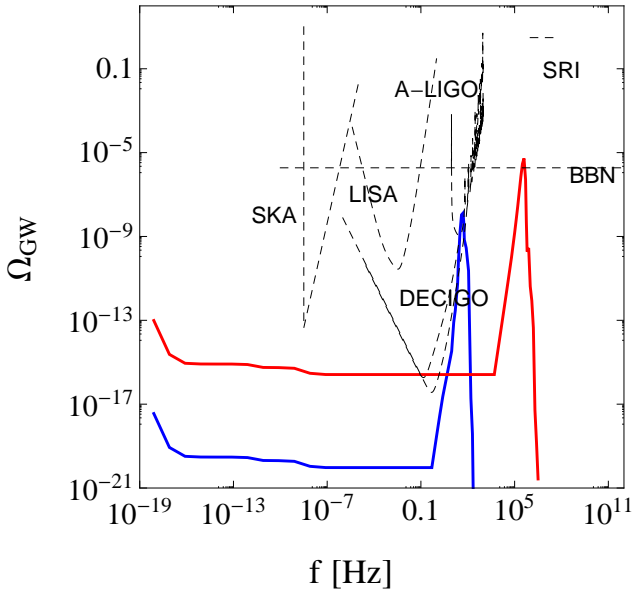}
\end{center}
\caption{\label{fig8}
$\Omega_{\rm GW}$ versus $f$ in low-scale 
inflation with the $\dot{\phi}$-dependent tensor 
mass squared (\ref{mg2}). The sensitivity curves of GW 
experiments are the same as shown in Fig.~\ref{fig7}. 
Each line corresponds to the models with 
(i) $M=10^8$~GeV, 
$\mu=1.523 \times 10^4$, $k_1=0.067M$, 
$t_\Gamma=10^{3}/M$ (red), and 
(ii) $M=1$~GeV, 
$\mu=4.78 \times 10^4$, $k_1=0.067M$, 
$t_\Gamma=10^{3}/M$ (blue).}
\end{figure}

In Fig.~\ref{fig8}, we plot today's GW background 
spectrum $\Omega_{\rm GW}(f,t_0)$ 
for $M=10^8$ GeV and $\Gamma=10^{-3}M$ 
with the coupling $\mu=1.523 \times 10^4$ (red line),
which are the same model parameters as 
those for Figs.~\ref{fig5} and \ref{fig6}.
For $H_i\simeq 2M/3$ as estimated in Eq.~(\ref{Hi}), 
the frequency formula (\ref{fi}) gives $f_i \simeq 5 \times 10^4$ Hz.
Since the peak wavenumber of ${\cal P}_T(k)$ 
shown in Fig.~\ref{fig6} is $k \simeq 5k_i$, the associated frequency 
at the maximum of $\Omega_{\rm GW}(f,t_0)$ is given by 
$f={\cal O}(10^5)$~Hz.
In this case, the peak amplitude is marginally consistent 
with the BBN bound.
The peak position is outside the detection range of the current 
ground-based GW measurements, but the predicted amplitude
reaches the sensitivity curve of DECIGO around the frequency 
$f=0.1$ Hz. 

There is a proposed detector design called 
the Synchronous Recycling Interferometer (SRI) \cite{SRI}
aiming to detect GWs at high frequencies around 
$10^6~{\rm Hz}<f<10^8~{\rm Hz}$. 
If the SRI reaches the sensitivity region below the BBN 
bound plotted in Fig.~\ref{fig8}, then we have the possibility 
to test our massive gravity scenario in 
low-scale inflation with 
$10^8~{\rm GeV} \lesssim M \lesssim 10^{12}~{\rm GeV}$.

In Fig.~\ref{fig8}, we also show $\Omega_{\rm GW}(f,t_0)$ 
for $M=1$ GeV and $\Gamma=10^{-3}M$ with the coupling 
$\mu=4.78 \times 10^4$ (blue line). 
In this case, the frequency formula (\ref{fi}) gives 
$f_i \simeq 5$~Hz with the peak wavenumber $k\simeq 5k_i$. 
This means that the spectrum is peaked at the frequency $f={\cal O}(10)$ Hz. 
For the coupling $\mu$ used in Fig.~\ref{fig8},
the maximum value of $\Omega_{\rm GW}(f,t_0)$ 
is $1.3 \times 10^{-8}$ at $f=50.2$ Hz, which 
satisfies the latest A-LIGO bound $\Omega_{\rm GW}(f,t_0)<1.7 \times 10^{-7}$ 
in the band $20~{\rm Hz}<f<86~{\rm Hz}$ \cite{ALIGO2}. 
This case is within the sensitivity region of the 
A-LIGO measurement. Thus there is a possibility for detecting 
the primordial GWs in the near future. 
The theoretical line is also on the verge of the DECIGO sensitivity curve. 
For increasing $\mu$ the amplitude of $\Omega_{\rm GW}(f,t_0)$ gets larger, 
so the present A-LIGO measurement places the upper bound 
$\mu \lesssim 5 \times 10^4$ for $M=1$~GeV.
Thus, in low-scale inflation with $M={\cal O}(1)$ GeV, 
it is possible to probe the physics of the massive tensor 
modes in direct GW measurements.

\section{Conclusions}
\label{consec}

In this paper, we studied the signature of the 
massive gravity theory in which the oscillating
tensor mass during reheating after inflation
gives rise to the parametric amplification 
of the primordial tensor perturbation.
For the theory described by the action 
(\ref{action}), the vector modes do not propagate
due to the internal symmetry (\ref{rescale}).
As a result, we are left with one scalar and two tensor 
propagating degrees of freedom with a time-dependent 
tensor mass. We identify the scalar degree of freedom 
as an inflaton field.

For the broad parametric resonance to occur during reheating, 
the tensor mass $m_g$ needs to be much larger than 
the Hubble expansion rate $H$. On the other hand, 
we require the condition $m_g^2 \ll H^2$ to avoid 
the strong suppression of massive GWs in the preceding 
inflationary epoch. 
This was made possible in our Lorentz-violating theory
because it is free from the Higuchi bound, $m_g^2>2H^2$~\cite{Higuchi:1986py}.
We proposed two explicit forms of 
$m_g^2$ satisfying these two requirements, see 
Eqs.~(\ref{mg1}) and (\ref{mg2}).
These requirements imply the existence of a transition from the regime 
$m_g^2 \lesssim H^2$ to the regime $m_g^2 \gtrsim H^2$ 
around the end of inflation. 

In Sec.~\ref{prisec}, we analytically estimated the tensor 
power spectrum ${\cal P}_T(k)$ at onset of reheating
by considering the transition of $m_g^2$ during inflation.
In Sec.~\ref{parasec}, we derived conditions for 
the occurrence of parametric resonance and typical
wavenumbers $k$ associated with the amplification of GWs.

In Sec.~\ref{Stasec}, we considered the Starobinsky model 
with the $\phi$-dependent tensor mass squared (\ref{mg1}) and
 numerically computed ${\cal P}_T(k)$ both at the onset and end of 
the amplification stage. As we see in Fig.~\ref{fig2}, the large-scale GWs
relevant to CMB observations have a nearly scale-invariant 
spectrum at the end of inflation, whereas the small-scale 
modes which were inside the Hubble radius during reheating 
have a highly blue-tilted spectrum. 
The parametric resonance leads to the amplification of GWs 
up to a cutoff wavenumber,  Eq.~(\ref{kcutcon}) (see Fig.~\ref{fig3}). 
The resulting power spectrum ${\cal P}_T(k)$, which 
is plotted in Fig.~\ref{fig4}, has a sharp peak around 
the wavenumber $k={\cal O}(10)k_i$.
For the modes satisfying the condition $k^2/a^2<m_g^2$, 
the amplitude of GWs decreases as 
$\langle {\cal P}_T \rangle \propto t^{-1}$ after reaching 
its maximum. This decrease continues until the time $t_{\Gamma}$ at which 
the inflaton decays to radiation. For $t>t_{\Gamma}$, the tensor perturbation 
behaves as in the standard massless case. Thus the final amplitude of the 
GW spectrum varies with the time $t_{\Gamma}$. 
The later the time $t_\Gamma$ is,
the smaller the amplitude becomes.

In Sec.~\ref{lowsec}, we studied low-scale inflation with the 
$\dot{\phi}$-dependent tensor mass squared (\ref{mg2}).
We considered a scenario in which the slow-roll parameter 
$\epsilon$ relevant to the CMB observations is much smaller 
than $10^{-4}$ and assumed a rapid transition to the 
potential  $V(\phi)=M^2\phi^2/2$, Eq.~(\ref{massivepo}),
to occur around $\phi \approx M_{\rm pl}$.
In such models, the power spectrum ${\cal P}_T$ at the beginning of reheating is very small, but the parametric resonance driven by 
the massive tensor can amplify GWs to the detectable 
level of CMB observations (see Fig.~\ref{fig5} for the case 
$M=10^8$ GeV). This mechanism is at work even for 
very low-scale inflation with the mass like $M=1$~GeV. 
The GWs can be efficiently amplified up to the wavenumber $k_{\rm cut}$
of order $10k_i$ where ${\cal P}_T(k, \tau_{\Gamma})$ is peaked
(see Fig.~\ref{fig6}).

We note that, in a low-scale inflationary scenario 
where the transition to the potential $V(\phi)=M^2\phi^2/2$ 
occurs for $\phi$ much smaller than $M_{\rm pl}$, 
the parametric resonance tends to be less efficient 
relative to the model studied in our paper for
the same coupling constant $\mu$. On the other hand, if we
consider a much larger $\mu$, one may be able to make the
parametric resonance efficient again. However, this will
rather generally lead to extremely strong suppression during
inflation unless one fine-tunes the behavior of $\dot\phi^2$ 
near the end of inflation. Such a case may be possible 
in models with a waterfall transition, but it is beyond
the scope of the present paper.

In Sec.\,\ref{GWbackground}, we computed today's energy density 
spectrum $\Omega_{\rm GW}(f, t_0)$ of the GW background in both Starobinsky 
inflation and low-scale inflation. The peak frequency $f_i$ of 
$\Omega_{\rm GW}(f, t_0)$ is given by the formula~(\ref{fi}).
In the Starobinsky model, the peak is at around $f=10^9$~Hz, 
which is much larger than the frequencies relevant to 
the current direct GW measurements (see Fig.~\ref{fig7}).
There is a range of frequencies around $f=0.1$ Hz 
in which the massive gravity scenario in Starobinsky inflation 
can reach the sensitivity curves of DECIGO and upgraded DECIGO, but the BBN bound is quite tight to limit the 
significant amplification of GWs during preheating.

Before concluding the paper, let us recapitulate a couple of
intriguing possibilities. One is the case of very low-scale
inflation. Since $H_i$ and $\Gamma$ in low-scale inflation can be 
much smaller than those in Starobinsky inflation, the peak of
 $\Omega_{\rm GW}(f, t_0)$ may appear at a much lower frequency.
For the model parameters $M=1$~GeV and $\Gamma=10^{-3}M$,
the peak of $\Omega_{\rm GW}(f, t_0)$ reaches
the sensitivity curve of A-LIGO, see Fig.~\ref{fig8}. 

The other is the possibility of high frequency GW observations.
If a future high frequency ($10^6~{\rm Hz}<f<10^8~{\rm Hz}$) 
GW detector like SRI can improve the sensitivity below the BBN bound,
it will offer the possibility for testing our massive gravity scenario in 
low-scale inflation with $10^8~{\rm GeV} \lesssim M \lesssim 10^{12}~{\rm GeV}$.

In conclusion, what is really exciting is that the existence of the 
time-dependent tensor mass in the early Universe can be potentially 
probed not only by CMB measurements but also by direct GW measurements at many 
different frequencies. Thus our massive gravity model provides
an ideal target for the multi-frequency gravitational wave astronomy
in years to come.

\acknowledgments 
This work is supported in part by the MEXT KAKENHI 
No.\,15H05888.
SK is supported by the Career Development Project for
Researchers of Allied Universities, and by JSPS Grant-in-Aid for
 Scientific Research No.\,17K14282.  
CL is supported by JSPS postdoc fellowship for overseas researchers, 
and by JSPS Grant-in-Aid for Scientific Research No.\,15F15321.
ST is supported by JSPS Grant-in-Aid for
Scientific Research No.~24540286 
and MEXT KAKENHI Grant-in-Aid for 
Scientific Research on Innovative Areas 
``Cosmic Acceleration'' 
(No.\,15H05890). 



\begin{thebibliography}{99}

\bibitem{Sta80}
A.~A.~Starobinsky,
Phys.\ Lett.\ B {\bf 91}, 99 (1980).

\bibitem{oldinf}
R.~Brout, F.~Englert and E.~Gunzig,
  Annals Phys.\  {\bf 115}, 78 (1978);
D.~Kazanas,
Astrophys.\ J.\  {\bf 241} L59 (1980);
K.~Sato, Mon.\ Not.\ R.\ Astron.\ Soc. {\bf 195}, 467 (1981);
Phys.\ Lett.\ {\bf 99B}, 66 (1981);
A.~H.~Guth,
Phys.\ Rev.\ D {\bf 23}, 347 (1981).

\bibitem{MukChib}
V.~F.~Mukhanov and G.~V.~Chibisov,
JETP Lett.\  {\bf 33}, 532 (1981).

\bibitem{oldper}
A.~H.~Guth and S.~Y.~Pi,
Phys.\ Rev.\ Lett.\  {\bf 49} (1982) 1110;
S.~W.~Hawking,
Phys.\ Lett.\ B {\bf 115}, 295 (1982);
A.~A.~Starobinsky,
Phys.\ Lett.\ B {\bf 117} (1982) 175;
J.~M.~Bardeen, P.~J.~Steinhardt and M.~S.~Turner,
Phys.\ Rev.\ D {\bf 28}, 679 (1983).

\bibitem{KS}
H.~Kodama and M.~Sasaki,
Prog.\ Theor.\ Phys.\ Suppl.\  {\bf 78}, 1 (1984);
V.~F.~Mukhanov,
JETP Lett.\  {\bf 41}, 493 (1985)
[Pisma Zh.\ Eksp.\ Teor.\ Fiz.\  {\bf 41}, 402 (1985)];
M.~Sasaki,
Prog.\ Theor.\ Phys.\  {\bf 76}, 1036 (1986);
V.~F.~Mukhanov, H.~A.~Feldman and R.~H.~Brandenberger,
Phys.\ Rept.\  {\bf 215}, 203 (1992).

\bibitem{Planckinf} 
P.~A.~R.~Ade {\it et al.} [Planck Collaboration],
Astron.\ Astrophys.\  {\bf 594}, A20 (2016)
[arXiv:1502.02114 [astro-ph.CO]].

\bibitem{GWsta} 
A.~A.~Starobinsky,
JETP Lett.\  {\bf 30}, 682 (1979)
[Pisma Zh.\ Eksp.\ Teor.\ Fiz.\  {\bf 30}, 719 (1979)].

\bibitem{review}
J.~E.~Lidsey, A.~R.~Liddle, E.~W.~Kolb, E.~J.~Copeland,
Rev.\ Mod.\ Phys.\  {\bf 69}, 373 (1997);
D.~H.~Lyth and A.~Riotto,
Phys.\ Rept.\  {\bf 314}, 1 (1999);
B.~A.~Bassett, S.~Tsujikawa and D.~Wands,
Rev.\ Mod.\ Phys.\  {\bf 78}, 537 (2006).

\bibitem{Lyth} 
D.~H.~Lyth,
Phys.\ Rev.\ Lett.\  {\bf 78}, 1861 (1997).

\bibitem{Bau} 
D.~Baumann and L.~McAllister,
Phys.\ Rev.\ D {\bf 75}, 123508 (2007)
[hep-th/0610285].

\bibitem{Tsuji14} 
S.~Tsujikawa,
PTEP {\bf 2014}, no. 6, 06B104 (2014)
[arXiv:1401.4688 [astro-ph.CO]].

\bibitem{DT10} 
A.~De Felice and S.~Tsujikawa,
Living Rev.\ Rel.\  {\bf 13}, 3 (2010)
[arXiv:1002.4928 [gr-qc]].

\bibitem{LiteBIRD}
M.~Hazumi {\it et al.} (LiteBIRD), 
Proc.\ SPIE Int.\ Soc.\ Opt.\ Eng.\ {\bf 8442}, 
844219 (2012).

\bibitem{stage4}
K.~N.~Abazajian {\it et al.} [CMB-S4 Collaboration],
arXiv:1610.02743 [astro-ph.CO].

\bibitem{sinflation} 
S.~Kachru, R.~Kallosh, A.~D.~Linde, J.~M.~Maldacena, 
L.~P.~McAllister and S.~P.~Trivedi,
JCAP {\bf 0310}, 013 (2003)
[hep-th/0308055];
D.~Baumann, A.~Dymarsky, I.~R.~Klebanov, J.~M.~Maldacena, 
L.~P.~McAllister and A.~Murugan,
JHEP {\bf 0611}, 031 (2006)
[hep-th/0607050];
S.~Panda, M.~Sami and S.~Tsujikawa,
Phys.\ Rev.\ D {\bf 76}, 103512 (2007)
[arXiv:0707.2848 [hep-th]];
D.~Baumann and L.~McAllister,
arXiv:1404.2601 [hep-th].

\bibitem{Lin15} 
C.~Lin and M.~Sasaki,
Phys.\ Lett.\ B {\bf 752}, 84 (2016)
[arXiv:1504.01373 [astro-ph.CO]].

\bibitem{GWpre} 
S.~Y.~Khlebnikov and I.~I.~Tkachev,
Phys.\ Rev.\ D {\bf 56}, 653 (1997)
[hep-ph/9701423];
J.~Garcia-Bellido and D.~G.~Figueroa,
Phys.\ Rev.\ Lett.\  {\bf 98}, 061302 (2007)
[astro-ph/0701014];
J.~Garcia-Bellido, D.~G.~Figueroa and A.~Sastre,
Phys.\ Rev.\ D {\bf 77}, 043517 (2008)
[arXiv:0707.0839 [hep-ph]].

\bibitem{Dubovsky1} 
S.~L.~Dubovsky,
JHEP {\bf 0410}, 076 (2004)
[hep-th/0409124].

\bibitem{Dubovsky2} 
S.~L.~Dubovsky, P.~G.~Tinyakov and I.~I.~Tkachev,
Phys.\ Rev.\ Lett.\  {\bf 94}, 181102 (2005)
[hep-th/0411158].

\bibitem{Fierz} 
M.~Fierz and W.~Pauli, Proc.\ Roy.\ Soc.\ Lond. {\bf A173}, 
211-232 (1939).

\bibitem{dRGT} 
C.~de Rham, G.~Gabadadze and A.~J.~Tolley,
Phys.\ Rev.\ Lett.\  {\bf 106}, 231101 (2011)
[arXiv:1011.1232 [hep-th]].

\bibitem{Higuchi:1986py} 
A.~Higuchi,
Nucl.\ Phys.\ B {\bf 282}, 397 (1987).

\bibitem{Domenech} 
G.~Domenech, T.~Hiramatsu, C.~Lin, M.~Sasaki, 
M.~Shiraishi and Y.~Wang,
JCAP {\bf 1705}, no. 05, 034 (2017)
[arXiv:1701.05554 [astro-ph.CO]].

\bibitem{KLS94} 
L.~Kofman, A.~D.~Linde and A.~A.~Starobinsky,
Phys.\ Rev.\ Lett.\  {\bf 73}, 3195 (1994)
[hep-th/9405187].

\bibitem{KLS97} 
L.~Kofman, A.~D.~Linde and A.~A.~Starobinsky,
Phys.\ Rev.\ D {\bf 56}, 3258 (1997)
[hep-ph/9704452].

\bibitem{ALIGO}
B.~P.~Abbott {\it et al.} [LIGO Scientific and Virgo Collaborations],
Phys.\ Rev.\ Lett.\  {\bf 116}, 061102 (2016)
[arXiv:1602.03837 [gr-qc]];
B.~P.~Abbott {\it et al.} [LIGO Scientific and Virgo Collaborations],
Phys.\ Rev.\ Lett.\  {\bf 116}, no. 24, 241103 (2016)
[arXiv:1606.04855 [gr-qc]].

\bibitem{DECIGO}
N.~Seto, S.~Kawamura and T.~Nakamura,
Phys.\ Rev.\ Lett.\  {\bf 87}, 221103 (2001)
[astro-ph/0108011];
S.~Kawamura {\it et al.},
Class.\ Quant.\ Grav.\  {\bf 23}, S125 (2006);
S.~Kawamura {\it et al.},
Class.\ Quant.\ Grav.\  {\bf 28}, 094011 (2011).

\bibitem{Labun} 
C.~Lin and L.~Z.~Labun,
JHEP {\bf 1603}, 128 (2016)
[arXiv:1501.07160 [hep-th]].

\bibitem{mpreheating} 
K.~Jedamzik and G.~Sigl,
Phys.\ Rev.\ D {\bf 61}, 023519 (2000)
[hep-ph/9906287];
P.~Ivanov,
Phys.\ Rev.\ D {\bf 61}, 023505 (2000)
[astro-ph/9906415];
A.~R.~Liddle, D.~H.~Lyth, K.~A.~Malik and D.~Wands,
Phys.\ Rev.\ D {\bf 61}, 103509 (2000)
[hep-ph/9912473];
C.~Gordon, D.~Wands, B.~A.~Bassett and R.~Maartens,
Phys.\ Rev.\ D {\bf 63}, 023506 (2001)
[astro-ph/0009131];
S.~Tsujikawa and B.~A.~Bassett,
Phys.\ Lett.\ B {\bf 536}, 9 (2002)
[astro-ph/0204031].

\bibitem{grabound} 
C.~de Rham, J.~T.~Deskins, A.~J.~Tolley and S.~Y.~Zhou,
Rev.\ Mod.\ Phys.\  {\bf 89}, no. 2, 025004 (2017)
[arXiv:1606.08462 [astro-ph.CO]].

\bibitem{chaotic} 
A.~D.~Linde,
Phys.\ Lett.\  {\bf 129B}, 177 (1983).

\bibitem{alpha} 
R.~Kallosh and A.~Linde,
JCAP {\bf 1310}, 033 (2013)
[arXiv:1307.7938 [hep-th]].

\bibitem{oldre}
A.~Albrecht, P.~J.~Steinhardt, M.~S.~Turner and F.~Wilczek,
Phys.\ Rev.\ Lett.\  {\bf 48}, 1437 (1982);
A.~D.~Dolgov and A.~D.~Linde,
Phys.\ Lett.\ B {\bf 116}, 329 (1982);
L.~F.~Abbott, E.~Farhi and M.~B.~Wise,
Phys.\ Lett.\ B {\bf 117}, 29 (1982).

\bibitem{Dolgov}
A.~D.~Dolgov and A.~D.~Linde,
Phys.\ Lett.\ B {\bf 116}, 329 (1982);
L.~F.~Abbott, E.~Farhi and M.~B.~Wise,
Phys.\ Lett.\ B {\bf 117}, 29 (1982).

\bibitem{Kuro14} 
S.~Kuroyanagi, S.~Tsujikawa, T.~Chiba and N.~Sugiyama,
Phys.\ Rev.\ D {\bf 90}, 063513 (2014)
[arXiv:1406.1369 [astro-ph.CO]].

\bibitem{Tsuji99} 
S.~Tsujikawa, K.~i.~Maeda and T.~Torii,
Phys.\ Rev.\ D {\bf 60}, 063515 (1999)
[hep-ph/9901306].

\bibitem{Kuro08} 
S.~Kuroyanagi, T.~Chiba and N.~Sugiyama,
Phys.\ Rev.\ D {\bf 79}, 103501 (2009)
[arXiv:0804.3249 [astro-ph]].

\bibitem{Allen} 
B.~Allen and J.~D.~Romano,
Phys.\ Rev.\ D {\bf 59}, 102001 (1999)
[gr-qc/9710117].

\bibitem{Maggiore} 
M.~Maggiore,
Phys.\ Rept.\  {\bf 331}, 283 (2000)
[gr-qc/9909001].

\bibitem{Yokoyama}
K.~Nakayama, S.~Saito, Y.~Suwa, and J.~Yokoyama,
Phys.\ Rev.\  {\bf D77}, 124001 (2008).
[arXiv:0802.2452 [hep-ph]];
JCAP {\bf 0806}, 020 (2008)
[arXiv:0804.1827 [astro-ph]].

\bibitem{Kuro11}
S.~Kuroyanagi, T.~Chiba and N.~Sugiyama,
Phys.\ Rev.\ D {\bf 83}, 043514 (2011)
[arXiv:1010.5246 [astro-ph.CO]].

\bibitem{Dai} 
L.~Dai, M.~Kamionkowski and J.~Wang,
Phys.\ Rev.\ Lett.\  {\bf 113}, 041302 (2014)
[arXiv:1404.6704 [astro-ph.CO]].

\bibitem{Yagi} 
K.~Yagi and N.~Seto,
Phys.\ Rev.\ D {\bf 83}, 044011 (2011)
[arXiv:1101.3940 [astro-ph.CO]].

\bibitem{LISA} 
P.~Amaro-Seoane {\it et al.},
Class.\ Quant.\ Grav.\  {\bf 29}, 124016 (2012)
[arXiv:1202.0839 [gr-qc]].

\bibitem{SKA} 
G.~Janssen {\it et al.},
PoS AASKA {\bf 14}, 037 (2015)
[arXiv:1501.00127 [astro-ph.IM]].

\bibitem{SRI} 
A.~Nishizawa {\it et al.},
Phys.\ Rev.\ D {\bf 77}, 022002 (2008)
[arXiv:0710.1944 [gr-qc]].
 A.~Nishizawa {\it et al.},
Class.\ Quant.\ Grav.\  {\bf 25}, 225011 (2008)
[arXiv:0801.4149 [gr-qc]].

\bibitem{Cabass:2015jwe}
G.~Cabass, L.~Pagano, L.~Salvati, M.~Gerbino, 
E.~Giusarma and A.~Melchiorri,
Phys.\ Rev.\ D {\bf 93}, 063508 (2016)
[arXiv:1511.05146 [astro-ph.CO]].
  
\bibitem{Cyburt:2015mya} 
R.~H.~Cyburt, B.~D.~Fields, K.~A.~Olive and T.~H.~Yeh,
Rev.\ Mod.\ Phys.\  {\bf 88}, 015004 (2016)
[arXiv:1505.01076 [astro-ph.CO]].

\bibitem{Salas} 
P.~F.~de Salas, S.~Gariazzo, J.~Lesgourgues and S.~Pastor,
JCAP {\bf 1709}, no. 09, 034 (2017)
[arXiv:1706.09850 [astro-ph.CO]].

\bibitem{Ade:2015xua} 
P.~A.~R.~Ade {\it et al.} [Planck Collaboration],
Astron.\ Astrophys.\  {\bf 594}, A13 (2016)
[arXiv:1502.01589 [astro-ph.CO]].

\bibitem{ALIGO2} 
B.~P.~Abbott {\it et al.} 
[LIGO Scientific and Virgo Collaborations],
Phys.\ Rev.\ Lett.\  {\bf 118}, no. 12, 121101 (2017)
[arXiv:1612.02029 [gr-qc]].

\end{thebibliography}
\end{document}